\documentclass[a4paper,11pt]{article}
\pdfoutput=1
\usepackage{float}
\usepackage{jcappub}
\usepackage{bm}
\usepackage{booktabs}
\usepackage{array}
\usepackage{microtype}

\newcommand{\dd}{\mathrm{d}}
\newcommand{\cR}{\mathcal{R}}
\newcommand{\cT}{\mathcal{T}}
\newcommand{\cH}{\mathcal{H}}
\newcommand{\OmGW}{\Omega_{\rm GW}}

\title{\boldmath Black Holes as Frequency-Dependent Filters of Stochastic Gravitational Waves}
\author[a]{Sefi Katznelson,}
\author[a]{Aidan Minger}
\author[b]{and Stefano Profumo}
\affiliation[a]{Department of Physics,
University of California, Santa Cruz, California 95064, USA}
\affiliation[b]{Department of Physics and Santa Cruz Institute for Particle Physics,\\
University of California, Santa Cruz, California 95064, USA}
\emailAdd{ykatznel@ucsc.edu}
\emailAdd{aminger@ucsc.edu}
\emailAdd{profumo@ucsc.edu}

\abstract{
Black holes ring when perturbed, whereas their response to a stationary
gravitational-wave background is a real-frequency scattering problem, not a
source of additional quasi-normal-mode lines.  We make this standard
distinction quantitative by treating a black hole as a frequency-, angle-, and
polarization-dependent filter.  For an isotropic stationary background around
Schwarzschild holes, elastic scattering produces no net monopole signal, so
horizon absorption is the only population-level spectral distortion.  We calculate this transfer function for
Schwarzschild holes in detail, identify its absorptive and phase-delay signatures, and
connect the stationary response to the causal ringdown excited by a finite
wave packet.  We then promote the single-hole result to an angular and
polarization transport kernel for a cosmological population.  The resulting
optical depth is negligible for realistic black-hole populations, including
asteroid-mass primordial black holes comprising all dark matter.  We then
extend the analysis to Kerr holes, for which superradiance allows genuine
amplification in selected co-rotating modes, but
isotropic incidence and random spin orientations strongly dilute the diffuse
signal.  Observable effects are therefore more likely in rare, nearby,
aligned, rapidly spinning, or transiently illuminated systems than through
cumulative cosmological propagation.}


\begin{document}
\maketitle
\flushbottom

\section{Introduction}
\label{sec:intro}

Black-hole quasi-normal modes are the characteristic poles of the retarded
response of a perturbed horizon.  Their complex frequencies are fixed by the
parameters of the background spacetime and govern the damped ringdown that
follows compact-binary coalescence or the scattering of a finite wave
packet~\cite{ReggeWheeler1957,Vishveshwara1970,Leaver1986,BertiReview2009}.
Excitation amplitudes are less universal than the frequencies: they depend on
the initial or boundary data, on the choice of perturbation variable, and on
the normalization used to define the residue of the Green function
pole~\cite{BertiCardoso2006,GlampedakisAndersson2003,OshitaCardoso2024,Lo2026}.

A conceptually distinct situation arises for a stationary stochastic
gravitational-wave background (SGWB).  After Fourier transformation, each real
frequency obeys a homogeneous Regge--Wheeler or Zerilli equation: the incident
GW enters as boundary data, and the outgoing wave at infinity is therefore the
reflected part of that same incident mode, not radiation generated by a new
source.  Time-translation invariance prevents frequency conversion, while the
QNM poles lie away from the real axis and shape the amplitude and phase of the
transfer function without injecting additional stationary power.  For a
Schwarzschild hole, the ingoing-horizon boundary condition and Wronskian
identity give $F_{\rm out}=F_{\rm in}-F_{\rm H}\leq F_{\rm in}$: any deficit in
the outgoing flux is energy absorbed by the horizon.  Elastic scattering only
redirects the remaining power and, for an isotropic background, scattering
into every line of sight exactly balances scattering out.  Consequently the
only net population-level change in the isotropic SGWB is absorption.  A
finite wave packet can still excite a decaying ringdown after the prompt
response has passed, but that is transient re-emission of previously supplied
energy, not a persistent QNM emission component.  Hawking radiation is a
separate quantum process, and Kerr superradiance is the separate case in which
rotational energy can supply genuine gain; neither is part of the passive
Schwarzschild signal considered here.

We do not present the absence of a stationary QNM line as a correction to an
established result of black-hole perturbation theory.  It follows directly
from linearity, time-translation invariance, and the analytic structure of the
retarded Green function.  The possible ambiguity is instead one of language:
``QNM excitation,'' excitation factors, and resonant response are often
discussed for initial-value problems or finite sources, and can be carried over
too literally to stationary illumination.  Our purpose is to put the transient
and stationary problems in a common normalization, identify which observable
features of the QNM poles survive on the real-frequency axis, and quantify the
resulting single-hole and population transfer.  Thus the contribution is a
rigorous quantitative framework and synthesis, not the resolution of an active
controversy.

These observations motivate a formulation directly in terms of measurable
scattering quantities, with Schwarzschild and Kerr playing complementary
roles.  For Schwarzschild holes we perform the detailed calculation: reflection
and horizon transmission as functions of $M\omega$, the complex phase and
finite-angle polarization response, causal wave-packet excitation, and the
resulting population transfer kernel.  The exact real-frequency amplitudes
provide a normalization-independent baseline, satisfy flux conservation, and
retain QNM physics through their analytic structure.  We then extend the
energy accounting to rotating holes, where selected co-rotating Teukolsky
modes are superradiantly amplified by extracting spin energy.  This Kerr part
is intentionally narrower: it uses the controlled low-frequency result and the
known strong-field modal maximum to establish the sign, scale, and population
dilution of the gain, rather than claiming a full Kerr analogue of the
Schwarzschild angular transfer calculation.

The claim tested in this paper is deliberately narrow.  A black hole is a
linear filter whose real-frequency transfer amplitudes encode both a
broad absorption transition and a sharper QNM-scale phase delay.  It does not,
under stationary driving, generate an independent line at a QNM frequency.
For a population, the same amplitudes define a radiative-transfer kernel and
the relevant expansion parameter is the optical depth
$\int n_{\rm BH}\sigma\,\dd s$.  The calculation therefore has two logically
distinct outcomes: a nontrivial and normalization-independent single-hole
response, and a restrictive population result showing that conventional
Schwarzschild populations are too optically thin to produce a large SGWB
distortion.  Rotation permits genuine gain in selected helicity and azimuthal
channels, but does not overcome population dilution for a statistically
isotropic background.

Three pieces of evidence establish this claim.  First, the real-frequency
reflection and transmission amplitudes obey flux conservation and display the
QNM scale without introducing a separate source term.  Second, their complex
phase, angular dependence, and polarization structure define a complete
finite-angle transfer kernel.  Third, integrating that kernel over a black-hole
population replaces nearest-neighbour estimates by a finite optical depth and
quantifies the smallness of the effect.  Results on primordial black holes and
Planck relics are applications of this framework, not separate claims.

The ingredients have substantial precedent, which fixes the scope of our
contribution.  Schwarzschild reflection, absorption, and helicity reversal are
standard results of black-hole scattering theory
\cite{Page1976,MatznerRyan1977,Dolan2008Long,Dolan2008}; QNM--greybody
relations make explicit that the same potential barrier controls poles and
real-frequency transmission~\cite{KonoplyaZhidenko2024}.  Pizzuti et al.
derived coupled Boltzmann equations for SGWB Stokes parameters in the
low-energy gravitational-Compton limit, including the spin-$4$ polarization
selection rule and the smallness of scattering by cosmological compact-object
populations~\cite{Pizzuti2023}.  Li, Hou, and Zhao subsequently reconstructed
finite-distance Schwarzschild-scattered waveforms and their physical
polarizations beyond the asymptotic forward-axis treatment
\cite{LiHouZhao2025}.  We do not claim a new scattering formalism or the first
spin-$4$ hierarchy.  Our contribution is to connect these strands at
$M\omega={\cal O}(1)$: we compute the absorptive, parity-resolved complex transfer
kernel, isolate its reflected-channel phase delay near the fundamental QNM,
project that finite-frequency kernel into angular and polarization collision
eigenmodes, and propagate the finite absorption through cosmological mass
functions.  A final Kerr extension places maximal superradiant gain beside the
same population scale and states the helicity and orientation conditions
required for amplification.

Section~\ref{sec:scattering} defines the boundary-value problem and the role of
QNM poles, demonstrates causal wave-packet excitation, and introduces Kerr
superradiance and an illustrative strong-field encounter;
Secs.~\ref{sec:sgwb} and~\ref{sec:numerics} construct and calculate the
stationary Schwarzschild transfer function.  Section~\ref{sec:population}
promotes it to a population kernel and optical depth.  We close by comparing
passive Schwarzschild filtering with mode-selective Kerr amplification.

\section{Black-hole scattering and QNM poles}
\label{sec:scattering}

\subsection{Regge--Wheeler and Zerilli problems}

We work in geometrized units $G=c=1$ and initially set the Schwarzschild mass
$M=1$.  Odd-parity gravitational perturbations obey
\begin{equation}
 \left[\frac{\dd^2}{\dd r_*^2}+\omega^2-V_\ell(r)\right]
 X_{\ell m\omega}(r)=0,
 \label{eq:rw}
\end{equation}
where
\begin{align}
 r_*&=r+2M\ln\left(\frac{r}{2M}-1\right),\\
 V_\ell(r)&=\left(1-\frac{2M}{r}\right)
 \left[\frac{\ell(\ell+1)}{r^2}-\frac{6M}{r^3}\right].
 \label{eq:potential}
\end{align}
Even-parity perturbations obey an equation of the same form for a master
variable $Z_{\ell m\omega}$, but with the Zerilli potential
\begin{equation}
 V_\ell^{Z}(r)=\left(1-\frac{2M}{r}\right)
 \frac{2\lambda^2(\lambda+1)r^3+6\lambda^2Mr^2
 +18\lambda M^2r+18M^3}
 {r^3(\lambda r+3M)^2},
 \qquad \lambda=\frac{(\ell-1)(\ell+2)}{2}.
 \label{eq:zerillipotential}
\end{equation}
The Regge--Wheeler and Zerilli potentials are Darboux partners: their
Schwarzschild transmission probabilities are equal, while their reflection
phases differ by the Chandrasekhar relation~\cite{Chandrasekhar1983}.  We do
not impose that relation numerically.  For the polarization-resolved angular
kernel, the odd- and even-parity equations are integrated independently from
$r=2M(1+10^{-6})$, in each case starting with the purely ingoing solution
$e^{-i\omega r_*}$.  At large radius we match the numerical solution and its
$r_*$ derivative to incoming and outgoing bases
$e^{\mp i\omega r_*}\sum_{n=0}^{N}a_n^\mp r^{-n}$, whose coefficients are
generated recursively from the corresponding potential.  We use $N=10$,
$r_{\max}/M=\max(800,2\ell^2)$, the \texttt{DOP853} integrator, and relative and
absolute tolerances $2\times10^{-10}$ and $2\times10^{-12}$.  Amplitudes are
extracted at 20 radii spanning $0.82r_{\max}$ to $r_{\max}$ and averaged.  The
independently recovered odd- and even-parity reflection probabilities agree to
$3.3\times10^{-7}$ or better; their unequal phases are retained when forming
the helicity-preserving and helicity-reversing amplitudes.

For real $\omega$, define the solution that is ingoing at the horizon by
\begin{align}
 X^{\rm in}_{\ell\omega}&\longrightarrow
 e^{-i\omega r_*}+\cR_\ell(\omega)e^{+i\omega r_*},
 &&r_*\rightarrow+\infty,\label{eq:bcinfinity}\\
 X^{\rm in}_{\ell\omega}&\longrightarrow
 \cT_\ell(\omega)e^{-i\omega r_*},
 &&r_*\rightarrow-\infty.\label{eq:bchorizon}
\end{align}
The incident wave has unit amplitude at infinity.  Conservation of the
Wronskian gives
\begin{equation}
 |\cR_\ell|^2+|\cT_\ell|^2=1,
 \label{eq:flux}
\end{equation}
so $|\cR_\ell|^2$ and $|\cT_\ell|^2$ are respectively reflection and
horizon-absorption probabilities in this normalization.  More explicitly,
$\cT_\ell$ is the amplitude of the purely ingoing horizon solution in
Eq.~\eqref{eq:bchorizon}, while $|\cT_\ell|^2$ is the fraction of the incident
energy flux that crosses the future horizon and is swallowed by the hole.  The
simple squared-amplitude relation follows because the asymptotic horizon and
infinity solutions have the same $|\omega|$ flux normalization; it should not
be read as transmission into a second observable asymptotic region.

Only $M\omega$ appears after rescaling all lengths by $M$.  Any dimensionless
response curve must consequently be independent of $M$ when plotted against
$M\omega$.  Mass changes the conversion to physical frequency but cannot
produce a family of different dimensionless curves.

\subsection{QNMs as poles, not additional stationary sources}

Let $X^{\rm up}$ denote the independent solution that is outgoing at infinity.
The radial retarded Green function is schematically
\begin{equation}
 G_\ell(r_*,r_*';\omega)=
 \frac{X^{\rm in}_\ell(r_<;\omega)X^{\rm up}_\ell(r_>;\omega)}
 {W_\ell(\omega)},
 \label{eq:green}
\end{equation}
where $W_\ell$ is the Wronskian.  QNM frequencies are zeros of
$W_\ell(\omega)$ in the lower half-plane.  Near a simple pole, a scattering
amplitude has the form
\begin{equation}
 \cR_\ell(\omega)=\cR_\ell^{\rm reg}(\omega)
 +\frac{B_{\ell n}}{\omega-\omega_{\ell n}}+\cdots.
 \label{eq:pole}
\end{equation}
The residue $B_{\ell n}$ is an excitation factor.  Its dimension depends on the
perturbation variable and normalization; only a complete observable assembled
from the residue and the physical boundary or source data is invariant
\cite{BertiCardoso2006,Lo2026}.

The inverse Fourier transform of Eq.~\eqref{eq:green} contains QNM residues,
prompt propagation, and a branch-cut contribution associated with late-time
tails~\cite{Leaver1986,CasalsOttewill2013}.  A pole-only sum is therefore neither
a complete Green function nor, by itself, a real-frequency scattering
observable.

For stationary driving, linear time-translation invariance gives
\begin{equation}
 \widetilde h_{\rm out}(f)=\cH(f)\widetilde h_{\rm in}(f).
 \label{eq:filter}
\end{equation}
The output frequency is the input frequency.  QNM poles can produce rapid
frequency dependence and phase evolution in $\cH$, while a freely damped
$e^{-i\omega_{\ell n}t}$ contribution describes a transient associated with the
initial data or switching history.

\subsection{Causal wave-packet excitation and outgoing ringdown}
\label{subsec:wavepacket}

The absence of a new stationary emission line does not preclude transient
emission.  To demonstrate the distinction directly, we solve the time-domain
Regge--Wheeler equation
\begin{equation}
 \left(\partial_t^2-\partial_{r_*}^2+V_2\right)X(t,r_*)=0
 \label{eq:rwtime}
\end{equation}
for an initially ingoing Gaussian packet,
\begin{equation}
 X(0,r_*)=e^{-(r_*-r_{*0})^2/(2\sigma_p^2)}
 \cos[\omega_0(r_*-r_{*0})],\qquad
 \partial_tX(0,r_*)=\partial_{r_*}X(0,r_*).
 \label{eq:packetinitial}
\end{equation}
The sign follows from the free-wave form $F(r_*+t)$: an ingoing packet moves
toward decreasing $r_*$ and therefore satisfies
$\partial_tF=\partial_{r_*}F$ at $t=0$.  We use $r_{*0}=140M$,
$\sigma_p=8M$, an extraction point $r_*=100M$, and
Sommerfeld boundaries at $r_*=-220M$ and $300M$.  A centered leapfrog scheme
with $\Delta r_*=0.1M$ and $\Delta t=0.05M$ evolves three packets with
$M\omega_0=0.30$, $0.37367$, and $0.60$.

\begin{figure}[t]
 \centering
 \includegraphics[width=0.88\linewidth]{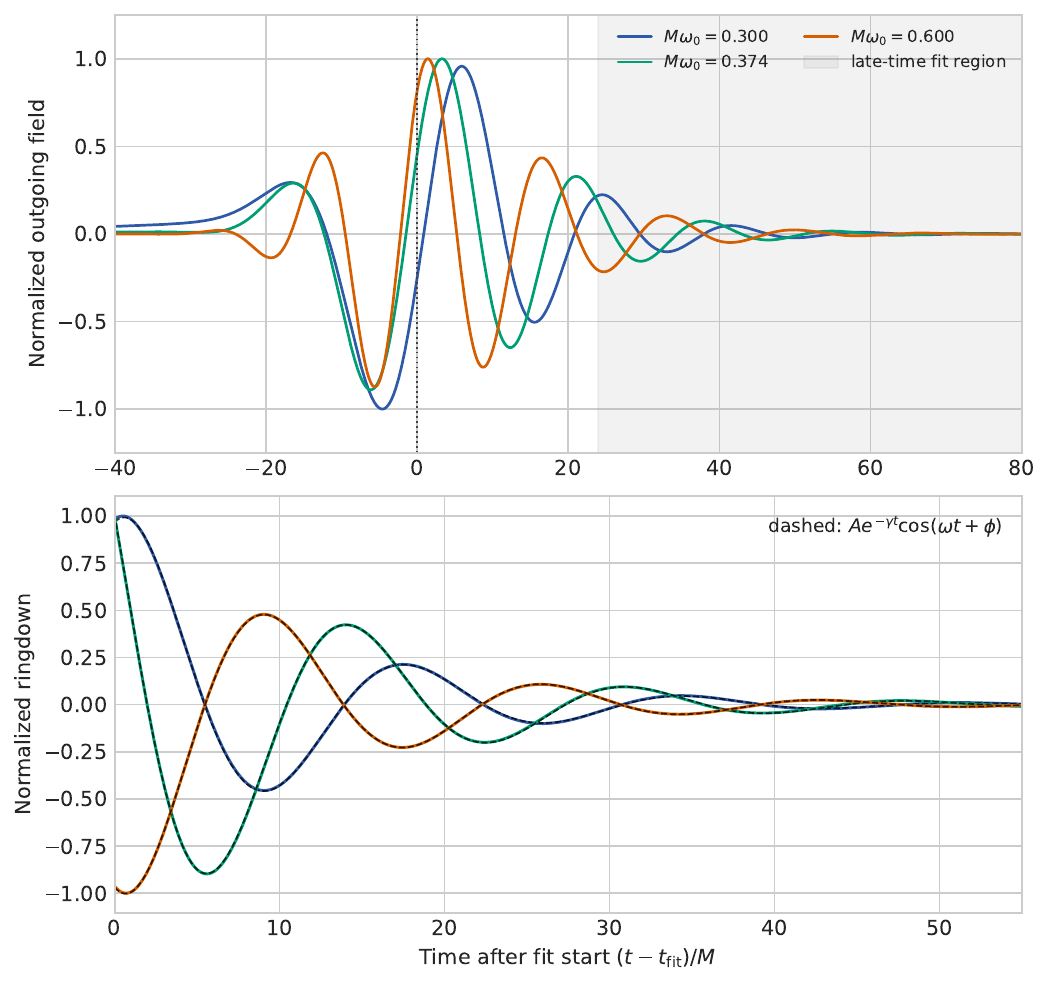}
 \caption{Causal scattering of finite quadrupolar packets (the x-axis is the same in both panels).  Upper: outgoing
 waveforms aligned on their prompt maxima and normalized separately; the
 shaded interval begins $3\sigma_p$ after each maximum.  Lower: late-time waveforms
 and damped-sinusoid fits (black dashed curves).  Despite different incident
 carrier frequencies, the late response approaches the fundamental
 Schwarzschild QNM.}
 \label{fig:wavepacket}
\end{figure}

Figure~\ref{fig:wavepacket} separates the prompt reflection from the subsequent
ringdown.  Beginning $3\sigma_p$ after the prompt maximum, we fit the extracted field
over $55M$ to $A e^{-\gamma t}\cos(\omega t+\phi)+C$.  The three fitted pairs
$(M\omega,M\gamma)$ are
\begin{equation}
 (0.3715,0.0912),\qquad(0.3717,0.0884),\qquad(0.3744,0.0883),
 \label{eq:wavepacketfits}
\end{equation}
to be compared with the fundamental value $(0.37367,0.08896)$.  This reference
is the Schwarzschild $\ell=2$, $n=0$ gravitational QNM in the
$e^{-i\omega t}$ convention, rounded from
$M\omega_{20}=0.37367168-0.08896232i$ as obtained by Leaver's
continued-fraction method and tabulated in the QNM compilation of
Refs.~\cite{Leaver1986,BertiReview2009}.  Varying the
fit start from $2.5\sigma_p$ to $3.5\sigma_p$ and its duration from $40M$ to $70M$ gives
$M\omega=0.3711$--$0.3729$ and $M\gamma=0.0880$--$0.0892$ for the central
packet.  A nested-grid test at fixed Courant ratio $\Delta t/\Delta r_*=0.5$
gives an observed waveform-convergence order $p=2.001$, as expected for the
centered leapfrog update; Appendix~\ref{app:tdconvergence} gives the definition
and numerical details.  Across those grids the fitted frequency and damping
rate remain stable at the few-$10^{-5}$ level, below the fit-window systematic.
Repeating the central-carrier calculation with
$\sigma_p/M=(6,8,10)$ gives fitted frequencies
$M\omega=(0.37012,0.37171,0.37239)$ and damping rates
$M\gamma=(0.08850,0.08841,0.08836)$.  The modest frequency drift reflects
changing prompt contamination, while the common damped response remains
consistent with the fundamental mode.

The energy current at the extraction point is
$F=-\partial_tX\,\partial_{r_*}X$.  The total reflected-energy fractions are
$0.624$, $0.359$, and $7.63\times10^{-3}$ for increasing carrier frequency,
consistent with the stationary filter transition.  The energy crossing the
extraction point after the adopted ringdown start is respectively
$1.22\times10^{-2}$, $1.02\times10^{-2}$, and $8.15\times10^{-5}$ of the
incident energy.  This late-time fraction is an operational measure, not a
unique QNM energy: prompt response, QNMs, and the power-law tail do not define
orthogonal components.  Nevertheless, the common fitted complex frequency
shows unambiguously that finite-duration illumination produces an outgoing
QNM transient.  Its energy comes from the incident packet; a Schwarzschild
hole supplies no stationary gain.

\subsection{Kerr superradiance: mode-selective gain}
\label{subsec:kerr}

Rotation changes the energy accounting.  For a Kerr hole with
$\chi=a/M$, horizons $r_\pm=M(1\pm\sqrt{1-\chi^2})$, and angular velocity
$\Omega_H=a/(2Mr_+)$, separated Teukolsky modes depend on
$e^{-i\omega t+im\phi}$~\cite{Teukolsky1972}.  Define their energy
amplification by
\begin{equation}
 Z_{s\ell m}=\frac{F_{\rm out}}{F_{\rm in}}-1.
 \label{eq:kerrgain}
\end{equation}
The horizon flux is negative, and hence $Z_{s\ell m}>0$, in the
superradiant interval
\begin{equation}
 0<\omega<m\Omega_H .
 \label{eq:superradiantcondition}
\end{equation}
The extra outgoing energy is rotational energy extracted from the hole.
Unlike Schwarzschild ringdown after a finite packet, this is genuine
stationary gain at the incident real frequency.

\begin{figure}[t]
 \centering
 \includegraphics[width=0.94\linewidth]{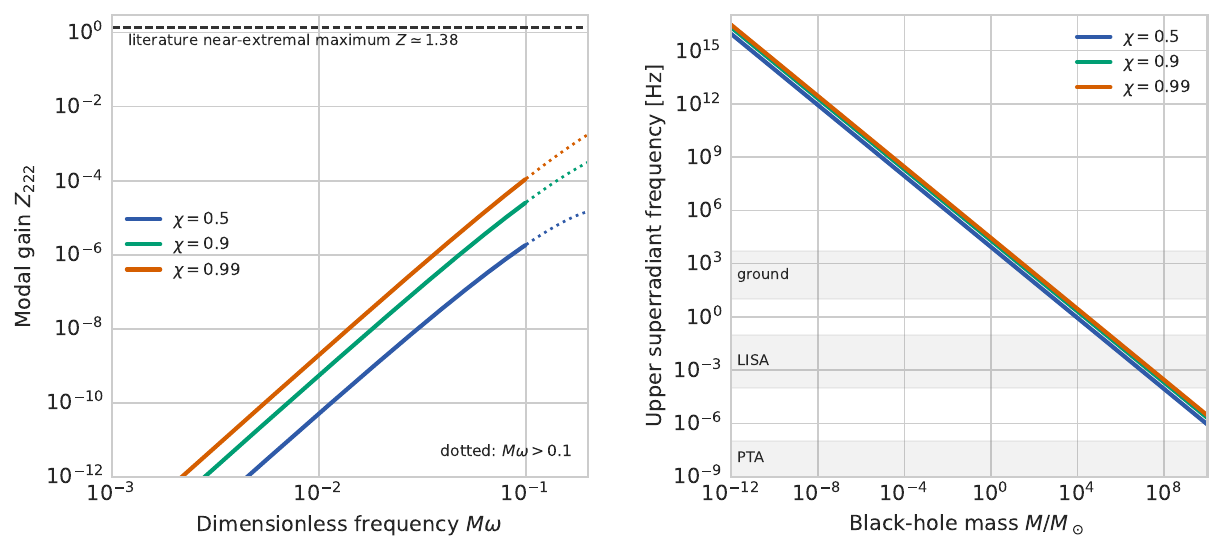}
 \caption{Kerr gravitational-wave superradiance.  Left: matched-asymptotic
 low-frequency modal gain for $s=\ell=m=2$; dotted curves lie beyond the
 conservative $M\omega\leq0.1$ domain.  The dashed line is the separate
 numerical literature maximum, not an extrapolation of these curves.  Right:
 upper edge $m\Omega_H/(2\pi)$ of the superradiant band versus mass, with
 indicative PTA, LISA, and ground-based frequency bands.}
 \label{fig:kerr}
\end{figure}

Figure~\ref{fig:kerr} shows two controlled parts of this result.  The left
panel evaluates the matched-asymptotic low-frequency expression for the
$s=\ell=m=2$ mode~\cite{TeukolskyPress1974,BritoCardosoPani2020}.  Solid
segments are restricted to $M\omega\leq0.1$; dotted segments indicate where
the approximation is becoming qualitative.  We do not use it to estimate the
strong-field peak.  Direct numerical integration of the Teukolsky equations
gives a maximum $Z_{222}\simeq1.38$ for a nearly extremal hole immediately
below the upper edge of the superradiant band; this is the gravitational
$s=\ell=m=2$ curve reported explicitly in Sec.~4.7.5 and Fig.~13 of
Ref.~\cite{BritoCardosoPani2020}.  We use Teukolsky and Press
\cite{TeukolskyPress1974} for the analytic low-frequency result, not as the
source of this strong-field numerical maximum.  The
right panel converts the exact kinematic boundary $m\Omega_H$ to physical
frequency and shows which masses can place this channel in standard detector
bands.

The modal maximum is not a sky-averaged cross section.  It selects the
co-rotating circular channel; the opposite helicity is absorbed, and a
linearly polarized plane gravitational wave has a net absorptive response
after both circular components are included~\cite{BritoCardosoPani2020}.
An isotropic, unpolarized SGWB incident on randomly oriented spins therefore
does not inherit a $138\%$ intensity gain.  Ensemble amplification requires a
correlation among wave helicity, propagation direction, and spin.

For aligned incidence there is a physically defined plane-wave gain cross
section.  In the conventions of Ref.~\cite{Dolan2008}, the on-axis circular
wave decomposes into spin-weighted spheroidal harmonics and the contribution of
one superradiant mode is
\begin{equation}
 \sigma_{\ell m}^{\rm gain}(\omega)
 =\frac{4\pi^2}{\omega^2}
 \left|{}_{-2}S_{\ell m}(0;a\omega)\right|^2 Z_{2\ell m}(\omega),
 \label{eq:kerrcrosssection}
\end{equation}
with the spheroidal harmonics normalized to unit angular integral.  This
formula replaces the earlier photon-capture-area construction: the factor
$27\pi M^2$ is a high-frequency Schwarzschild absorption limit and has no
special connection to the superradiant $(2,2)$ mode.  Equation
\eqref{eq:kerrcrosssection} also makes clear that a modal amplification factor
alone does not determine a cross section; the incident-wave projection and
frequency are essential.

Dolan's direct numerical on-axis calculation gives a peak negative total
absorption cross section of approximately $-6M^2$ for a co-rotating circular
wave at $a=0.99M$ and $M\omega\simeq0.8$ (Fig.~3 of
Ref.~\cite{Dolan2008}).  We therefore use
$\sigma_{\rm gain}^{\rm axis}\simeq6M^2$ as a concrete computed benchmark,
not as a near-extremal upper bound.  Inserted into the same cosmic-mean
one-Hubble-path estimate used below, it gives
\begin{equation}
 n\sigma_{\rm gain}^{\rm axis}\frac{c}{H_0}
 =\frac{f_{\rm BH}\Omega_{\rm DM}\rho_{\rm crit}}{M}
 6\left(\frac{GM}{c^2}\right)^2\frac{c}{H_0}
 =\frac{9\Omega_{\rm DM}}{4\pi}f_{\rm BH}
 \frac{H_0GM}{c^3},
 \label{eq:kerrtausubstitution}
\end{equation}
where $\rho_{\rm crit}=3H_0^2/(8\pi G)$.  With
$\Omega_{\rm DM}=0.264$, $H_0=67.4\,\mathrm{km\,s^{-1}\,Mpc^{-1}}$, and
$GM_\odot/c^3=4.9255\,\mu\mathrm{s}$, this evaluates to
\begin{equation}
 \tau_{\rm gain}^{\rm axis}\simeq2.0\times10^{-24}f_{\rm BH}
 \left(\frac{M}{M_\odot}\right),
 \label{eq:kerrtau}
\end{equation}
before helicity and spin-orientation suppression.  For an asteroid-mass PBH
with $M=10^{-16}M_\odot$ this is $2.0\times10^{-40}$ even if such objects are
all of the dark matter.  Kerr superradiance reverses the sign of the
single-channel energy exchange without rescuing cumulative propagation
through a homogeneous population.

An intentionally favorable source--hole encounter can evade the homogeneous
population suppression, but estimating it requires additional geometric and
coherence assumptions.  We therefore relegate the corresponding schematic
construction and Fig.~\ref{fig:rareencounter} to
Appendix~\ref{app:rareencounter}; it is not used in any population bound.

\section{Stochastic incident radiation}
\label{sec:sgwb}

Decompose the incident field into propagation direction $\hat{\bm n}$ and
polarization $A$.  We define a one-sided covariance by
\begin{equation}
 \left\langle\widetilde h_A(f,\hat{\bm n})
 \widetilde h_{A'}^*(f',\hat{\bm n}')\right\rangle
 =\frac12\delta(f-f')\delta_{AA'}
 \delta^2(\hat{\bm n},\hat{\bm n}')\,\mathcal I(f),
 \label{eq:covariance}
\end{equation}
where conventional angular normalization factors can be moved between
$\mathcal I$ and the delta function.  Statistical isotropy makes the covariance
diagonal in angular momentum and independent of $m$; it does not restrict the
field to even $\ell$.  Because a spin-2 field has $\ell\geq2$, the black-hole
response begins with the quadrupole.

For a fixed incident and observed angular channel, Eq.~\eqref{eq:filter} implies
\begin{equation}
 S_{\rm out}(f)=|\cH(f)|^2S_{\rm in}(f).
 \label{eq:psdtransfer}
\end{equation}
If the detector receives both the incident and scattered fields, the measured
PSD is
\begin{equation}
 S_{\rm tot}=S_{\rm in}+S_{\rm sc}
 +2\operatorname{Re}S_{\rm in,sc}.
 \label{eq:interference}
\end{equation}
The cross spectrum is generally nonzero because the two fields arise from the
same stochastic realization.  It may vanish after specified angular,
population, or instrumental averaging, but this is an observable-dependent
statement rather than an automatic consequence of stochasticity.

At a large distance $r$ from one black hole, the scattered strain can be
written as
\begin{equation}
 \widetilde h_{\rm sc}(f,r,\hat{\bm x})=
 \frac{F(f,\hat{\bm x},\hat{\bm n})}{r}
 e^{2\pi ifr/c}\widetilde h_{\rm in}(f,\hat{\bm n}).
 \label{eq:farfield}
\end{equation}
The scattering amplitude $F$ has dimensions of length.  Schwarzschild scale
invariance requires
\begin{equation}
 F=M\widehat F(M\omega,\hat{\bm x},\hat{\bm n}),
\end{equation}
and hence
\begin{equation}
 S_{\rm sc}(f,r)=\frac{M^2}{r^2}
 |\widehat F(M\omega)|^2S_{\rm in}(f),
 \qquad
 \frac{\dd\sigma}{\dd\Omega}=M^2|\widehat F|^2.
 \label{eq:scaling}
\end{equation}
There is one propagation factor from the black hole to the observer.  An
additional $r^{-2}$ in the PSD would require the illuminating intensity at the
black hole to decrease with that same distance, which is not the geometry of a
spatially homogeneous SGWB.

\section{Numerical real-frequency response}
\label{sec:numerics}

\subsection{Method and validation}

We integrate Eq.~\eqref{eq:rw} in $r$, evolving the pair
$(X,P)$ with $P=\dd X/\dd r_*$:
\begin{equation}
 \frac{\dd X}{\dd r}=\frac{P}{1-2M/r},\qquad
 \frac{\dd P}{\dd r}=-\frac{\omega^2-V_\ell}{1-2M/r}X.
 \label{eq:firstorder}
\end{equation}
The inner boundary is $r=2M(1+10^{-6})$, with the ingoing asymptotic solution
$X=e^{-i\omega r_*}$ and $P=-i\omega X$.  At large radius the numerical
solution is decomposed into $e^{\mp i\omega r_*}$ over multiple fitting radii.
We use an eighth-order explicit Runge--Kutta integrator with relative and
absolute tolerances $2\times10^{-10}$ and $2\times10^{-12}$.  The outer boundary
is chosen as $r_{\rm max}/M=\max[250,80/(M\omega)]$.

The principal internal validation is Eq.~\eqref{eq:flux}.  Across the complete
grid $0.035\leq M\omega\leq1.2$ and $2\leq\ell\leq12$, the maximum absolute
flux defect is $1.3\times10^{-5}$; most of the grid is substantially more accurate.
The complete tabulated response and the script used to generate it are included
in the supplementary material described in the Data Availability Statement.
Table~\ref{tab:uncertaintybudget} collects the numerical
stability tests used throughout the analysis and distinguishes them from
model-dependent variations such as the background adopted in the descriptive
delay fit.

All numerical products were generated with Python 3.12.13, NumPy 2.3.5,
SciPy 1.17.0, and Matplotlib 3.10.8.  The stationary radial equations use
SciPy's \texttt{solve\_ivp} implementation of the explicit eighth-order
\texttt{DOP853} scheme.  The time-domain packet uses the centered leapfrog
update described in Sec.~\ref{subsec:wavepacket}.  The population calculation
uses a shape-preserving \texttt{PchipInterpolator} for
$\widehat\sigma_{\rm abs}$ and composite trapezoidal integration on logarithmic
mass and redshift grids.  Angular moments use composite Simpson integration;
Legendre and Jacobi polynomials are evaluated with \texttt{scipy.special}.
The Wigner-$d$ elements are implemented explicitly in the Condon--Shortley
convention as their factorial--Jacobi representation and checked against the
closed $L=4$ forms.  The pole audit uses \texttt{scipy.interpolate.AAA}.

\subsection{Results}

\begin{figure}[t]
 \centering
 \includegraphics[width=0.88\linewidth]{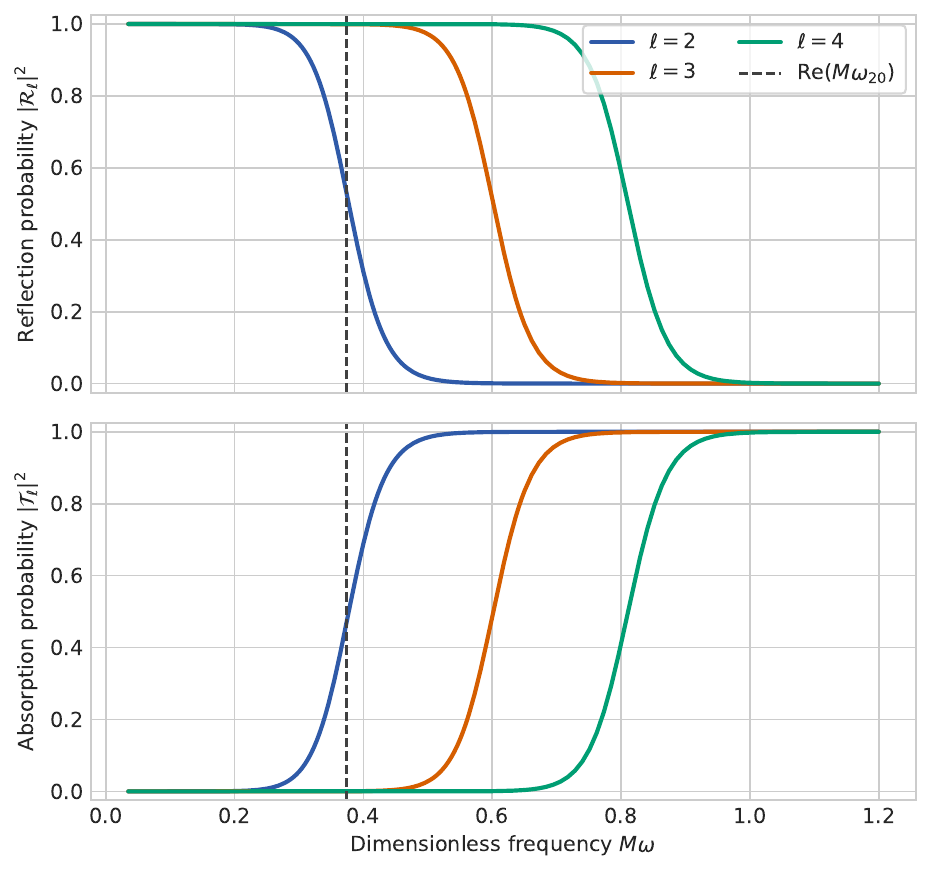}
 \caption{Reflection (upper panel) and horizon-absorption (lower panel)
 probabilities for axial gravitational perturbations of a Schwarzschild black
 hole.  The dashed line marks the real part of the fundamental quadrupolar QNM
 frequency.  The QNM scale lies within a broad $\ell=2$ filtering transition;
 higher multipoles are strongly suppressed there.}
 \label{fig:response}
\end{figure}

Figure~\ref{fig:response} shows the exact reflection and absorption
probabilities.  At low frequency the curvature potential reflects almost all
incident radiation.  Each multipole turns from reflection to absorption as the
frequency crosses its potential barrier, with higher multipoles turning over
at larger $M\omega$.

For the fundamental Schwarzschild quadrupole, using the literature reference
specified in Sec.~\ref{subsec:wavepacket},
\begin{equation}
 M\omega_{20}=0.37367-0.08896i.
\end{equation}
At its real part, our numerical results give
\begin{equation}
 |\cT_2|^2\simeq0.469,\qquad
 |\cT_3|^2\simeq2.37\times10^{-4},\qquad
 |\cT_4|^2\simeq6.7\times10^{-8}.
 \label{eq:numericalvalues}
\end{equation}
Thus the quadrupolar QNM scale lies close to the midpoint of the $\ell=2$
filter transition.  It does not appear as a narrow peak in a positive-definite
absorption probability.  The pole is nevertheless part of the analytic
structure controlling the response, and may be more sharply exposed through
the complex phase, time delay, or a transient waveform.

\subsection{Complex reflection phase and Wigner delay}
\label{subsec:delay}

The QNM-scale structure is clearer in the phase than in the reflection probability.
This complements established QNM--greybody correspondences based on
transmission probabilities~\cite{KonoplyaZhidenko2024} by retaining the phase
of the reflected gravitational channel.
Define the partial-wave reflection $S$ matrix by
\begin{equation}
 S_\ell(\omega)=(-1)^{\ell+1}\cR_\ell(\omega)
 =\eta_\ell(\omega)e^{2i\delta_\ell(\omega)},
 \label{eq:smatrix}
\end{equation}
where $0\leq\eta_\ell\leq1$ accounts for horizon absorption.  The group delay
of the reflected channel is~\cite{Wigner1955,Smith1960}
\begin{equation}
 \tau_\ell(\omega)=2\frac{\dd\operatorname{Re}\delta_\ell}{\dd\omega}
 =\frac{\dd}{\dd\omega}\arg S_\ell(\omega).
 \label{eq:wigner}
\end{equation}
Because the reflected channel is not unitary by itself, Eq.~\eqref{eq:wigner}
is the delay conditional on reflection, not the trace of the full unitary
Wigner--Smith matrix.  Its dimensional scaling is $\tau_\ell=M\widehat
\tau_\ell(M\omega)$.

\begin{figure}[t]
 \centering
 \includegraphics[width=0.88\linewidth]{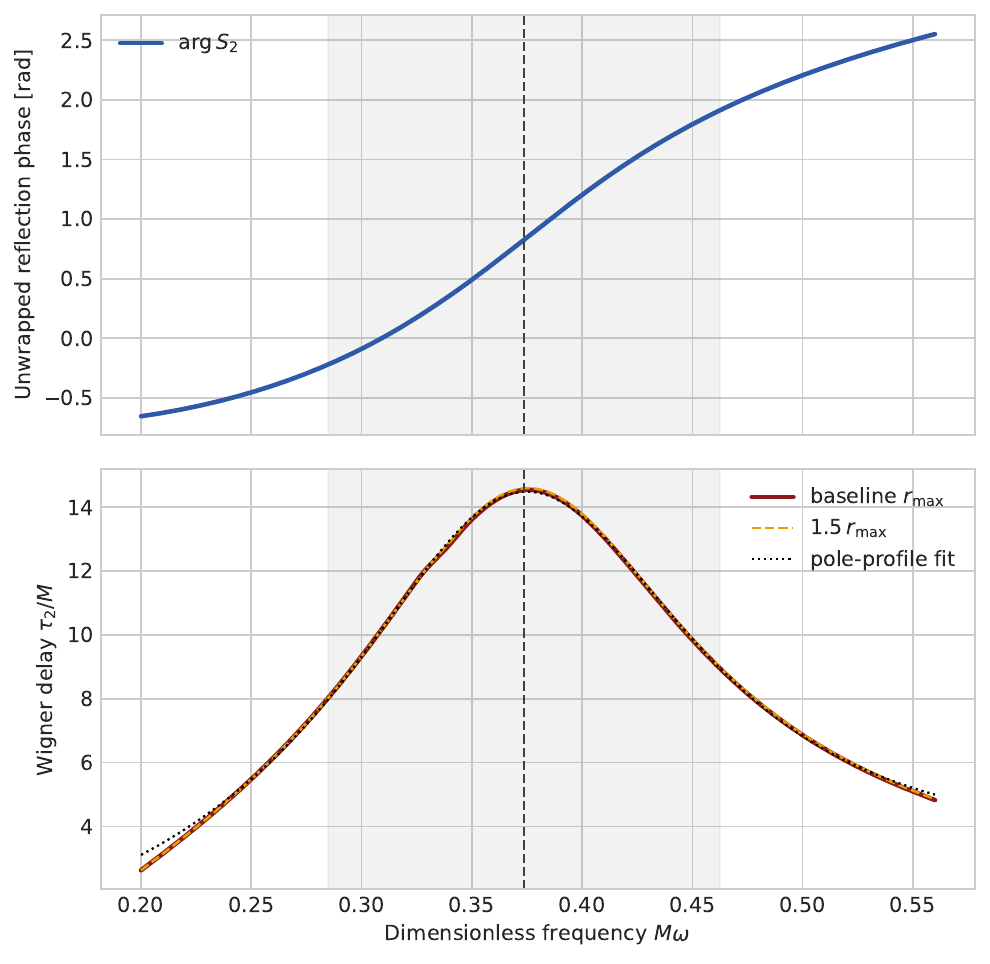}
 \caption{Unwrapped phase of the quadrupolar reflection $S$ matrix (upper
 panel) and reflected-channel Wigner delay (lower panel).  The vertical dashed
 line marks $\operatorname{Re}(M\omega_{20})$ and the gray band extends one
 $|\operatorname{Im}(M\omega_{20})|$ to either side.  The orange dashed curve
 repeats the calculation with an extraction radius $50\%$ larger.  The dotted
 curve is the descriptive pole-profile fit of Eq.~\eqref{eq:delayfit}.  The
 localized delay maximum identifies the QNM scale more clearly than the
 reflection probability; Eq.~\eqref{eq:aaapole} supplies the separate rational
 pole extraction.}
 \label{fig:delay}
\end{figure}

This conditional delay nevertheless has an operational meaning.  For a
coherent, narrow-band incident packet prepared in a definite partial wave,
stationary-phase propagation shifts the centroid of the reflected packet by
$\tau_\ell$ relative to a reference propagation phase, provided that the
amplitude and phase vary slowly across the packet bandwidth.  It could
therefore be inferred from a phase-resolved comparison of an incident waveform
with its separately identified reflected echo.  The qualification
``conditional'' is essential: the reflected signal power is additionally
weighted by $|\cR_\ell|^2$, the delay becomes ill-conditioned near a
reflection zero, and an inclusive dwell time requires the complete multichannel
Wigner--Smith operator.  A stationary stochastic background supplies neither a
common phase reference nor, for an unresolved population, an identifiable
incident/reflected pair.  Consequently we do not treat the peak in
Fig.~\ref{fig:delay} as a directly observable SGWB time lag.  In this paper it
is primarily a phase-sensitive diagnostic of the QNM-scale analytic structure;
the adaptive
Antoulas--Anderson (AAA) barycentric rational algorithm
\cite{NakatsukasaSeteTrefethen2018} continuation below provides the actual pole-location cross-check.

We calculate Eq.~\eqref{eq:wigner} on a uniform grid of 241 frequencies over
$0.20\leq M\omega\leq0.56$.  The unwrapped phase is locally smoothed with a
third-order Savitzky--Golay polynomial over 17 grid points before taking its
analytic derivative.  Figure~\ref{fig:delay} shows a pronounced delay feature:
\begin{equation}
 \frac{\tau_{2,\rm max}}{M}=14.54,
 \qquad M\omega_{\rm max}=0.3755.
 \label{eq:delaypeak}
\end{equation}
At $M\omega=\operatorname{Re}(M\omega_{20})$, the result is
$\tau_2/M=14.54$ to the displayed precision.  In physical units,
\begin{equation}
 \tau_{2,\rm max}\simeq71.6\,\mu\mathrm{s}
 \left(\frac{M}{M_\odot}\right).
 \label{eq:delayphysical}
\end{equation}

To quantify the association with the pole, we fit the delay over
$0.23<M\omega<0.54$ to a Lorentzian pole profile plus a locally linear
background,
\begin{equation}
 \frac{\tau_2}{M}=c_0+c_1(x-x_c)
 +A\frac{\Gamma}{(x-x_c)^2+\Gamma^2},\qquad x=M\omega.
 \label{eq:delayfit}
\end{equation}
For the fiducial fit the result is
\begin{equation}
 x_c=0.3726,\qquad \Gamma=0.1009,
 \label{eq:delayfitvalues}
\end{equation}
with an rms residual $0.042M$.  The fitted center is close to
$\operatorname{Re}(M\omega_{20})=0.37367$, while the width is about $13\%$
larger than $|\operatorname{Im}(M\omega_{20})|=0.08896$.  Varying the
Savitzky--Golay window from 9 to 31 points and its polynomial order from 2 to 5
gives $0.3755$ for the peak position and $14.39$--$14.55$ for its height.
Varying the fit interval and allowing either a linear or quadratic smooth
background gives $x_c=0.3718$--$0.3736$ and
$\Gamma=0.086$--$0.107$.  Thus the center is robustly QNM-scale, but agreement
of the fitted width with the damping rate is only at the 10--20\% level.  The
delay fit is descriptive rather than an independent QNM determination: regular
scattering, absorption, and the analytic zero structure of the reflection
amplitude also contribute to the phase, and a genuine pole extraction requires
analytic continuation.

We perform that continuation directly as a cross-check.  The complex
$\cR_2(\omega)$ values on the real axis are approximated by the AAA
\cite{NakatsukasaSeteTrefethen2018}.  Poles of the rational approximant provide
an analytic continuation of the same scattering amplitude whose poles coincide
with those of the retarded Green function.  Without using the reference QNM to
select the answer, we retain the largest-residue pole in the broad box
$0.25<\operatorname{Re}(M\omega)<0.50$ and
$-0.20<\operatorname{Im}(M\omega)<-0.03$.  Across 60 combinations of five
frequency windows, four sampling densities, and three approximation tolerances,
the median pole is
\begin{equation}
 M\omega_{20}^{\rm AAA}=0.37389-0.08932i.
 \label{eq:aaapole}
\end{equation}
The full ranges are $0.37317$--$0.37437$ for the real part and
$0.08829$--$0.09010$ for $-\operatorname{Im}(M\omega)$.  Every extraction is
therefore within $0.19\%$ of the reference oscillation frequency and $1.29\%$
of its damping rate.  The residue magnitude remains $0.0945$--$0.1026$, and
the nearest pole--zero separation exceeds $0.045$, excluding a near-canceling
Froissart doublet.  Appendix~\ref{app:aaapole} gives the stability plot and
selection details.  This is an actual pole extraction from the numerical
real-axis amplitude, although it is a cross-check of the same scattering
solution rather than an independent perturbation code.

Increasing the outer extraction radius by $50\%$ changes the delay by at most
$0.067M$ over the region where $|\cR_2|^2>0.03$ and leaves the peak location
unchanged at the grid resolution.  The flux-conservation defect on this denser
calculation is below $2.1\times10^{-6}$.  A separate calculation across the
peak with twice the extraction radius moves the maximum by at most one grid
interval and changes its height by $0.08M$; tightening the integration
tolerances and moving the inner boundary from $r/2M-1=10^{-6}$ to $10^{-7}$
changes the height by $5\times10^{-5}M$.

The overall delay contains a convention-dependent smooth contribution: a shift
of the origin of $r_*$ changes the reflection phase by a term linear in
$\omega$.  We use the standard tortoise-coordinate convention of
Eq.~\eqref{eq:potential}.  The localized peak, its position relative to the QNM
frequency, and the agreement between extraction radii are insensitive to a
constant phase choice; comparisons of absolute delays should retain the same
asymptotic convention or subtract a smooth reference background.

The conversion to physical frequency is
\begin{equation}
 f_{220}=\frac{\operatorname{Re}(M\omega_{20})}{2\pi}
 \frac{c^3}{GM}
 \simeq12.1\,\mathrm{kHz}\left(\frac{M_\odot}{M}\right).
 \label{eq:massfrequency}
\end{equation}
Here $c^3/(GM_\odot)=2.03\times10^5\,\mathrm{s}^{-1}$, so the factor
$\operatorname{Re}(M\omega_{20})/(2\pi)$ gives $12.1\,\mathrm{kHz}$.  This
dimensional mapping is only a frequency correspondence, not a sensitivity
forecast.

\section{From one black hole to a population}
\label{sec:population}

\subsection{Three checks of the far-field scaling}

The scattered PSD scales as $r^{-2}$ for three independent reasons.  First, an
outgoing Schwarzschild master function has asymptotically constant radial
amplitude, while the radiative metric perturbation reconstructed at future null
infinity is proportional to $X/r$.  Thus $h_{\rm sc}\propto r^{-1}$ and its PSD
is proportional to $r^{-2}$.  Second, the plane-wave scattering definition
$\dd\sigma/\dd\Omega=r^2 F_{\rm sc}/F_{\rm in}$ requires the scattered flux to
fall as $r^{-2}$.  Third, the scattering amplitude in
Eq.~\eqref{eq:farfield} must have dimensions of length, so the only
Schwarzschild scale gives $F=M\widehat F(M\omega)$ and reproduces
Eq.~\eqref{eq:scaling}.  These statements agree with standard partial-wave
black-hole scattering calculations~\cite{MatznerRyan1977,Dolan2008}.

The $r^{-4}$ law can arise in a different geometry.  If a localized source is a
distance $d$ from the black hole and the black hole is a distance $r$ from the
observer, then
\begin{equation}
 h_{\rm sc}\sim h_{\rm src,0}\frac{1}{d}\frac{M}{r}\widehat F,
 \qquad
 S_{\rm sc}\propto\frac{M^2}{d^2r^2}S_{\rm src,0}.
 \label{eq:twodistance}
\end{equation}
Setting $d=r$ produces an $r^{-4}$ PSD, but this special source--hole--observer
configuration is not a homogeneous SGWB.  A background specified by its local
intensity at every black-hole position contains no additional observer-centered
$d^{-2}$ factor.

\subsection{Incoherent shell sum and radiative transfer}

For incoherent scattered intensities, a Euclidean shell contributes
\begin{equation}
 \dd S_{\rm sc}^{\rm pop}\sim
 n_{\rm BH}(4\pi r^2\dd r)\frac{M^2}{r^2}
 |\widehat F|^2S_{\rm in}
 =4\pi n_{\rm BH}M^2|\widehat F|^2S_{\rm in}\,\dd r.
 \label{eq:shell}
\end{equation}
The contribution is not nearest-neighbour dominated; a naive homogeneous
integral is sensitive to the available path length.  The natural dimensionless
quantity is instead an optical depth,
\begin{equation}
 \tau(f)=\int n_{\rm BH}(s)\sigma(f,s)\,\dd s.
 \label{eq:tau}
\end{equation}
For cosmological propagation,
\begin{equation}
 \tau(f_0)=\int_0^{z_{\rm max}}
 \frac{c\,\dd z}{(1+z)H(z)}n_{\rm BH}(z)
 \sigma\!\left[(1+z)f_0;M,z\right],
 \label{eq:taucosmological}
\end{equation}
where $n_{\rm BH}$ is the physical number density and the response is evaluated
at the rest-frame frequency.  A mass function is included by inserting
$\int\dd M\,\dd n_{\rm BH}/\dd M$.  Appendix~\ref{app:opticaldepthderivation}
derives this expression directly from the local attenuation law and shows how
the factors of $(1+z)$ enter.

More explicitly, suppressing polarization indices, the local collision term in
a radiative-transfer equation is
\begin{align}
 C[I](f,\hat{\bm n})={}&-n_{\rm BH}
 [\sigma_{\rm abs}(f)+\sigma_{\rm sc}(f)]I(f,\hat{\bm n})\nonumber\\
 &+n_{\rm BH}\int\dd\Omega'\,
 \frac{\dd\sigma(\hat{\bm n}'\!\to\!\hat{\bm n})}{\dd\Omega}
 I(f,\hat{\bm n}').
 \label{eq:collision}
\end{align}
For an isotropic field $I=I_0(f)$, rotational invariance and the definition of
$\sigma_{\rm sc}$ make the elastic loss and gain terms cancel.  The collision
term reduces to $C[I_0]=-n_{\rm BH}\sigma_{\rm abs}I_0$.  Elastic scattering
therefore cannot create or alter the monopole intensity of an exactly
homogeneous isotropic background.  It can redistribute anisotropies and
polarization.  A nonzero monopole source term requires absorption followed by a
specified emission process, spatially varying illumination, evolving sources,
finite boundaries, rotation with energy exchange, or other non-equilibrium
physics.

There is an additional infrared subtlety for the elastic term.  The
long-wavelength gravitational differential cross section behaves as
$\dd\sigma/\dd\Omega\propto M^2/\sin^4(\theta/2)$ in the forward direction
\cite{MatznerRyan1977,Dolan2008}.  Its angle-integrated total is therefore
forward divergent.  A population calculation based on a total elastic
scattering optical depth needs an angular-resolution or impact-parameter
cutoff, or else a coherent wave-propagation treatment in which very small-angle
deflections contribute primarily to phase and lensing.  The absorption cross
section is finite and provides the cleanest population observable.

\subsection{Finite-angle kernel and transport weighting}
\label{subsec:angularkernel}

For a circularly polarized plane wave incident on
a Schwarzschild black hole, write the elastic differential cross section as
\begin{equation}
 \frac{\dd\sigma}{\dd\Omega}=|f(\theta)|^2+|g(\theta)|^2,
 \label{eq:fgcrosssection}
\end{equation}
where $f$ preserves helicity and $g$ reverses it.  In the controlled
low-frequency limit, $M\omega\ll1$, the two contributions are
\begin{equation}
 \frac{|f|^2}{M^2}=\frac{\cos^8(\theta/2)}{\sin^4(\theta/2)},
 \qquad
 \frac{|g|^2}{M^2}=\sin^4(\theta/2)
 \label{eq:lowfrequencykernel}
\end{equation}
\cite{MatznerRyan1977,Dolan2008Long,Dolan2008}.  The first term gives
$\dd\sigma/\dd\Omega\sim16M^2/\theta^4$ as $\theta\to0$; the second is
finite and dominates the backward direction.  Thus the parity dependence of
the Schwarzschild phase shifts produces a nonzero helicity-reversing channel
even without rotation.

For a measurement that resolves only deflections larger than
$\theta_{\min}$, define
\begin{align}
 \sigma_{>}(\theta_{\min})&=2\pi\int_{\theta_{\min}}^\pi
 \sin\theta\,\frac{\dd\sigma}{\dd\Omega}\,\dd\theta,\label{eq:sigmaresolved}\\
 \sigma_{\rm tr,>} (\theta_{\min})&=2\pi\int_{\theta_{\min}}^\pi
 \sin\theta(1-\cos\theta)\frac{\dd\sigma}{\dd\Omega}\,\dd\theta.
 \label{eq:sigmatransport}
\end{align}
As the cutoff is reduced, $\sigma_{>}\propto\theta_{\min}^{-2}$ and
$\sigma_{\rm tr,>}\propto\ln(1/\theta_{\min})$.  A sharp $\theta_{\min}$ is only
a calculational proxy, not a universal detector angle.  For a particular SGWB
map or cross-correlation analysis it must be replaced by the detector-network
and map-making response.  Schematically, an intensity multipole is governed by

\begin{equation}
 \kappa_L^I[\mathcal R]=2\pi\int_0^\pi\!\dd\theta\,\sin\theta\,
 \mathcal R(\theta)\,[1-P_L(\cos\theta)]
 \frac{\dd\sigma}{\dd\Omega},
 \label{eq:responsewindow}
\end{equation}

where $\mathcal R(\theta)$ is the effective angular response of the analysis;
a hard cut corresponds to $\mathcal R=\Theta(\theta-\theta_{\min})$.  Angles
inside the unresolved forward cone primarily contribute coherently to phase
and lensing and should not be counted as independent local collisions.

We use $20^\circ$ in the finite-frequency plots because the twice-reduced
partial-wave reconstruction is explicitly converged at and above that angle
(Appendix~\ref{app:diagnostics}), not because $20^\circ$ is a detector beam
width.  The cutoff sensitivity can be displayed analytically in the
low-frequency limit:

\begin{table}[H]
\centering
\begin{tabular}{cccc}
\toprule
$\theta_{\min}$ & $\sigma_{\rm tr,>}/M^2$ & ratio to $20^\circ$
& rescaled dipole-depth proxy \\
\midrule
$1^\circ$  & 192.26 & 4.28 & $2.2\times10^{-22}\,f_{\rm BH}(M/M_\odot)$ \\
$5^\circ$  & 111.56 & 2.49 & $1.3\times10^{-22}\,f_{\rm BH}(M/M_\odot)$ \\
$10^\circ$ & 77.33  & 1.72 & $9.0\times10^{-23}\,f_{\rm BH}(M/M_\odot)$ \\
$20^\circ$ & 44.89  & 1.00 & $5.2\times10^{-23}\,f_{\rm BH}(M/M_\odot)$ \\
$30^\circ$ & 28.27  & 0.63 & $3.3\times10^{-23}\,f_{\rm BH}(M/M_\odot)$ \\
\bottomrule
\end{tabular}
\caption{Dependence of the low-frequency transport cross section on a hard
 forward-cone cutoff.  The last column applies the corresponding ratios to the
 finite-frequency, one-Hubble-length dipole-depth benchmark quoted below; it is
 a sensitivity proxy rather than a new finite-frequency calculation.  Even an
 aggressive $1^\circ$ choice changes the already negligible benchmark by only
 a factor of 4.28.  The isotropic absorption depth is independent of every
 entry in this table.}
\label{tab:cutoffsensitivity}
\end{table}

Figure~\ref{fig:angularkernel} shows the continuous low-frequency dependence.
The population conclusion based on the isotropic monopole is stronger still:
elastic gain and loss cancel exactly, and the remaining absorption optical
depth contains no angular cutoff at all.

At finite $M\omega$, Eqs.~\eqref{eq:fgcrosssection}--\eqref{eq:sigmatransport}
remain the appropriate definitions, but $f$ and $g$ require the independently
computed even- and odd-parity phase shifts.  We use the parity-resolved
integration and asymptotic-series matching specified below
Eq.~\eqref{eq:zerillipotential}; retaining $N=10$ terms removes the centrifugal
and curvature phase contamination present in a bare plane-wave extraction.  At
$M\omega=0.37367$, the flux defect remains below $2.3\times10^{-10}$; changing
the asymptotic order from 10 to 13 changes the $\ell=60$ phase by
$2.3\times10^{-9}$ radians.  The odd- and even-parity reflection probabilities
agree to $3.3\times10^{-7}$ or better, with the largest discrepancy in the
absorbing $\ell=2$ channel.

We then apply two Yennie-type reductions to the spin-weighted partial-wave
series~\cite{Dolan2008}, using phase shifts through $\ell=60$ and a physical
truncation at $\ell_{\max}=50$.  The convergence tests are collected in
Appendix~\ref{app:diagnostics}.  At the QNM
frequency the converged values of $M^{-2}\dd\sigma/\dd\Omega$ are 351.3,
43.25, 5.754, 0.5558, 0.7044, and 0.9590 at scattering angles
$30^\circ,60^\circ,90^\circ,120^\circ,150^\circ,$ and $180^\circ$,
respectively.

\begin{figure}[p]
 \centering
 \includegraphics[width=0.88\linewidth]{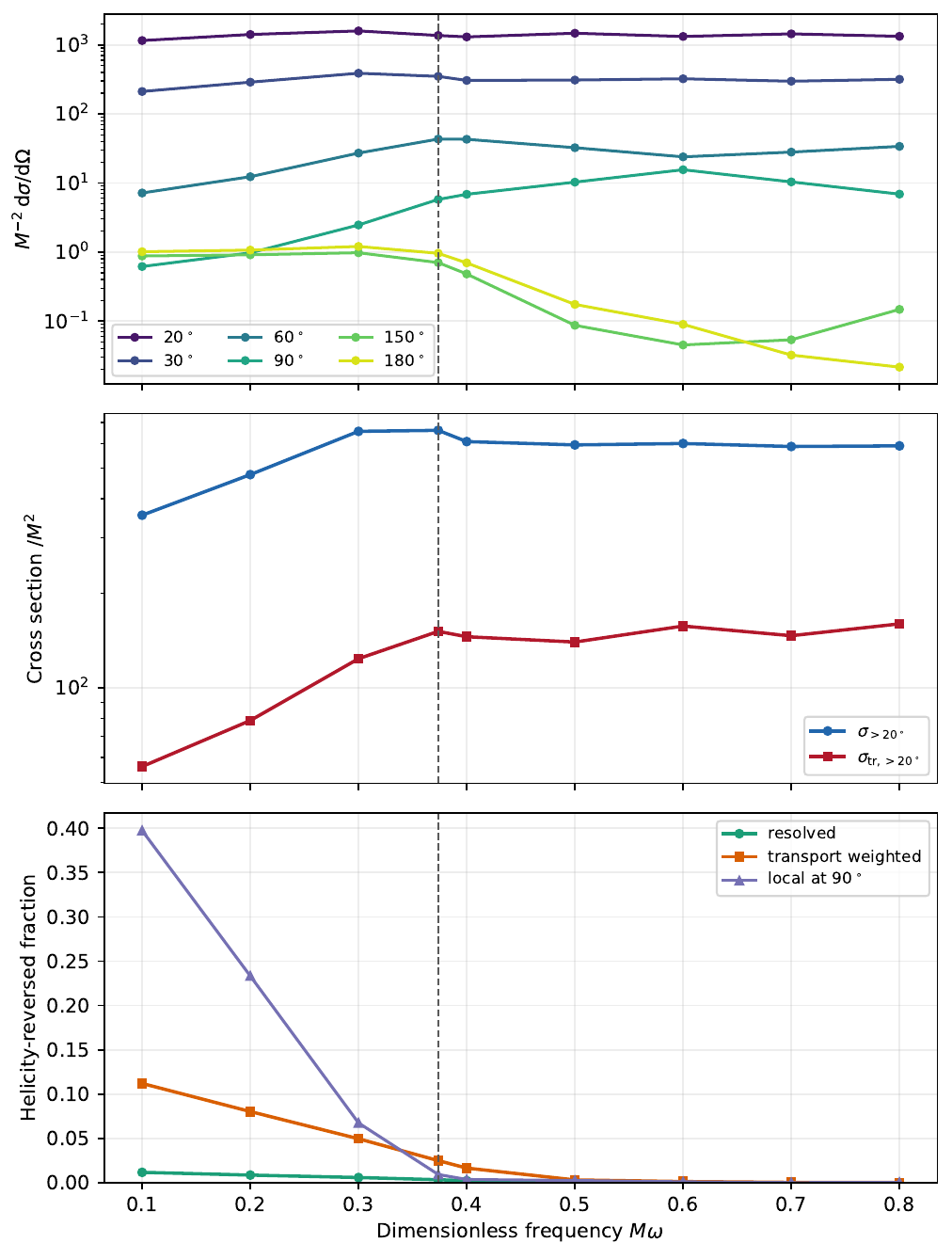}
 \caption{Finite-frequency Schwarzschild scattering over
 $0.1\leq M\omega\leq0.8$.  Top: differential cross section at representative
 finite angles.  Middle: resolved and transport-weighted cross sections above
 $20^\circ$.  Bottom: the helicity-reversed fraction of the resolved and
 transport-weighted signals, together with its local value at $90^\circ$.
 The dashed line marks the real part of the fundamental quadrupolar QNM
 frequency.}
 \label{fig:finitegrid}
\end{figure}

We extend the same reconstruction over nine frequencies spanning
$0.1\leq M\omega\leq0.8$.  Figure~\ref{fig:finitegrid} displays both local
angular kernels and finite-angle moments above the numerically verified
$20^\circ$ boundary.  The resolved cross section rises from
$354.6M^2$ at $M\omega=0.1$ to a broad maximum of $660.9M^2$ at the sampled
QNM frequency, then remains near $600M^2$.  The transport-weighted result is
$56.10M^2$ at $M\omega=0.1$ and $151.2M^2$ at $M\omega=0.37367$; its subsequent
oscillation between approximately $140M^2$ and $160M^2$ demonstrates that the
QNM scale is not a narrow resonance in the angle-integrated redistribution.

The parity-sensitive result is sharper.  The helicity-reversed fraction of
$\sigma_{>20^\circ}$ falls monotonically from $1.18\%$ at $M\omega=0.1$ to
$3.45\times10^{-3}$ at the QNM frequency and $4.97\times10^{-5}$ at
$M\omega=0.8$.  Transport weighting emphasizes large angles and therefore
raises these fractions to $11.2\%$, $2.50\%$, and $2.40\times10^{-4}$,
respectively, without changing the trend.  Across the grid, the 95th
percentile of the pointwise $\ell_{\max}=40$--50 difference is below
$6.9\times10^{-4}$.  One versus two series reductions can differ by several
percent at isolated angular interference minima, but the resolved and
transport moments change by at most $4.5\times10^{-3}$ and
$9.1\times10^{-4}$, respectively.  These integrated quantities are therefore
the more robust inputs for radiative transfer.

\clearpage

\subsection{Angular and polarization transfer}
\label{subsec:boltzmannmultipoles}

Following the SGWB Stokes/Boltzmann construction of
Ref.~\cite{Pizzuti2023}, the finite-angle kernel can be projected into the
collision operator for an isotropic population of Schwarzschild black holes.
The distinction here is that the collision kernel is obtained from the
finite-frequency black-hole amplitudes, rather than the low-energy
gravitational-Compton limit.  Let $I_+$ and $I_-$
denote the intensities of the two circular helicities and define
$I=I_++I_-$ and $V=I_+-I_-$.  With
$F(\theta)=|f(\theta)|^2$ and $G(\theta)=|g(\theta)|^2$, elastic scattering
preserves helicity through $F$ and reverses it through $G$.  Rotational
invariance makes each spherical-harmonic multipole an eigenmode of the
collision operator.  For a resolved angular interval
$\theta_{\min}\leq\theta\leq\pi$, the corresponding damping cross sections are
\begin{align}
 \kappa_L^I={}&2\pi\int_{\theta_{\min}}^\pi\!\dd\theta\,\sin\theta
 [1-P_L(\cos\theta)](F+G),\label{eq:kappaI}\\
 \kappa_L^V={}&2\pi\int_{\theta_{\min}}^\pi\!\dd\theta\,\sin\theta
 \{[1-P_L(\cos\theta)]F+[1+P_L(\cos\theta)]G\}.
 \label{eq:kappaV}
\end{align}
The collision terms are
$C[I_{LM}]=-n_{\rm BH}(\sigma_{\rm abs}+\kappa_L^I)I_{LM}$ and likewise for
$V_{LM}$ with $\kappa_L^V$.

Several nontrivial checks follow immediately.  Equation~\eqref{eq:kappaI}
gives $\kappa_0^I=0$ exactly, demonstrating elastic monopole conservation.
For $L=1$, $1-P_1=1-\cos\theta$, so that
$\kappa_1^I=\sigma_{\rm tr,>}$ exactly.  Circular polarization behaves
differently: $\kappa_0^V=2\sigma_{g,>}$ because every helicity-reversing event
changes the sign of $V$.  Thus an isotropic intensity is unaffected by elastic
scattering, whereas an isotropic circular-polarization asymmetry is
depolarized.

At $M\omega=0.37367$ and $\theta_{\min}=20^\circ$,
$\kappa_1^I=151.23M^2$ and $\kappa_0^V=4.556M^2$.  The corresponding
mean-density, one-Hubble-length dipole depth is
$5.2\times10^{-23}f_{\rm BH}(M/M_\odot)$, so angular projection does not alter
the population conclusion.  The full multipole curves are shown in
Appendix~\ref{app:diagnostics}.

The power kernels $F$ and $G$ are sufficient for $I$ and $V$ when the incident
linear polarization vanishes, but a complete polarization calculation must
retain the relative phase of $f$ and $g$.  In scattering-plane bases, the
helicity-space Jones matrix gives the Mueller map
\begin{equation}
 \begin{pmatrix}I'\\Q'\\U'\\V'\end{pmatrix}=
 \begin{pmatrix}
 S&A&0&0\\ A&S&0&0\\0&0&D&-B\\0&0&B&D
 \end{pmatrix}
 \begin{pmatrix}I\\Q\\U\\V\end{pmatrix},
 \quad
 \begin{matrix}
 S=|f|^2+|g|^2,&D=|f|^2-|g|^2,\\
 A=2\Re(fg^*),&B=2\Im(fg^*).
 \end{matrix}
 \label{eq:muellermatrix}
\end{equation}
An unpolarized incident beam acquires $Q/I=A/S$ in the scattering-plane basis,
while $B/S$ mixes $U$ and $V$.  Direct reconstruction of the physical
polarizations therefore requires both parity sectors and their relative
phases~\cite{LiHouZhao2025}.

For gravitational radiation, $Q\pm iU$ carry spin weight $\mp4$, rather than
the spin weight two familiar from electromagnetic Stokes fields
\cite{Pizzuti2023}.  We therefore
project Eq.~\eqref{eq:muellermatrix} with Wigner functions
$d^L_{40}$, $d^L_{44}$, and $d^L_{4,-4}$.  In the Condon--Shortley convention,
define
\begin{align}
 K_L^{IE}&=2\pi\int\dd\theta\,\sin\theta\,d^L_{40}(\theta)A(\theta),
 &K_L^{VB}&=2\pi\int\dd\theta\,\sin\theta\,d^L_{40}(\theta)B(\theta),
 \label{eq:spin4sources}\\
 K_L^{F}&=2\pi\int\dd\theta\,\sin\theta\,d^L_{44}(\theta)F(\theta),
 &K_L^{G}&=2\pi\int\dd\theta\,\sin\theta\,d^L_{4,-4}(\theta)G(\theta).
 \label{eq:spin4gains}
\end{align}
The resolved linear-polarization damping eigenvalues are then
\begin{equation}
 \kappa_L^E=\sigma_{>}-K_L^F-K_L^G,
 \qquad
 \kappa_L^B=\sigma_{>}-K_L^F+K_L^G,
 \qquad L\ge4.
 \label{eq:spin4damping}
\end{equation}
Parity symmetry makes the scalar source purely $I\to E$: there is no
$I\to B$ term.  The imaginary interference kernel instead supplies the
$V\to B$ coupling in Eq.~\eqref{eq:spin4sources}; reversing the polarization
basis changes the signs assigned to $U$, $B$, and this coefficient together,
without changing an observable prediction.

At $M\omega=0.37367$ and $\theta_{\min}=20^\circ$, the $L=4$ results are
\begin{equation}
 (\kappa_4^E,\kappa_4^B,K_4^{IE},K_4^{VB})/M^2
 =(221.96,224.51,1.903,2.787).
 \label{eq:spin4numbers}
\end{equation}
The integrated $I\to E$ source is small compared with the damping eigenvalue;
the complete Mueller curves and convention checks appear in the appendices.

\subsection{Finite absorption depth and benchmark magnitude}

For an incident plane gravitational wave, the Schwarzschild absorption cross
section can be written in terms of the partial transmission probabilities as
\begin{equation}
 \sigma_{\rm abs}(\omega)=\frac{\pi}{\omega^2}
 \sum_{\ell=2}^{\infty}(2\ell+1)|\cT_\ell(\omega)|^2
 =M^2\widehat\sigma_{\rm abs}(M\omega)
 \label{eq:absorptioncrosssection}
\end{equation}
\cite{Page1976,Sanchez1978,Dolan2008}.  Using the numerical values in
Eq.~\eqref{eq:numericalvalues}, at the fundamental quadrupolar QNM scale we
obtain
\begin{equation}
 \widehat\sigma_{\rm abs}(0.37367)\simeq52.75,
 \label{eq:sigmares}
\end{equation}
with the $\ell=2$ term dominant.

For a monochromatic black-hole mass function, dimensional analysis gives an
absorption coefficient of order
\begin{equation}
 \alpha_{\rm abs}(f)=n_{\rm BH}\sigma_{\rm abs}(f)
 \sim \frac{\rho_{\rm BH}}{M}M^2\widehat\sigma_{\rm abs}(M\omega)
 =\rho_{\rm BH}M\widehat\sigma_{\rm abs}(M\omega).
 \label{eq:alphascaling}
\end{equation}
This linear dependence on path length and on $\rho_{\rm BH}M$ is qualitatively
different from a nearest-neighbour scaling proportional to
$\rho_{\rm BH}^{4/3}M^{-4/3}$.

Restoring SI units and allowing black holes to compose a fraction $f_{\rm BH}$
of a mass density $\rho$, the QNM-scale estimate is
\begin{equation}
 \tau_{\rm abs}\simeq52.75\,f_{\rm BH}\rho
 \frac{G^2M}{c^4}L.
 \label{eq:tauestimate}
\end{equation}
For a solar-mass population comprising all of the mean cosmological dark matter
over one Hubble length, this gives approximately
\begin{equation}
 \tau_{\rm abs}\sim1.8\times10^{-23}f_{\rm BH}
 \left(\frac{M}{M_\odot}\right),
 \label{eq:taucbench}
\end{equation}
before including redshift evolution.  For a local density
$0.4\,\mathrm{GeV\,cm^{-3}}$ over $10\,\mathrm{kpc}$, the corresponding estimate
is $1.3\times10^{-23}f_{\rm BH}(M/M_\odot)$.  Frequency matching alone therefore
does not imply detectability.  Larger masses increase the depth at fixed mass
density, but their QNM feature moves to lower frequency and their allowed
cosmic density must be evaluated for the specific population.

\subsection{Cosmological mass-function convolution}
\label{subsec:cosmologicaldepth}

We now evaluate Eq.~\eqref{eq:taucosmological} rather than replacing the
cosmological path by a single characteristic length.  Let $\psi(M)$ be the
black-hole mass fraction per logarithmic mass interval, normalized by
$\int\psi(M)\,\dd\ln M=1$.  If black holes comprise a fraction $f_{\rm BH}$ of
the present dark-matter density, their physical differential number density is
\begin{equation}
 \frac{\dd n_{\rm BH}}{\dd\ln M}=
 f_{\rm BH}\rho_{\rm DM,0}(1+z)^3\frac{\psi(M)}{M}.
 \label{eq:massfunctionnumber}
\end{equation}
The absorption depth observed at frequency $f_0$ is therefore
\begin{equation}
 \frac{\tau_{\rm abs}(f_0)}{f_{\rm BH}}=
 \int_0^{z_{\max}}\!\frac{c\,\dd z}{(1+z)H(z)}
 \rho_{\rm DM,0}(1+z)^3
 \int\!\dd\ln M\,\frac{\psi(M)}{M}
 \left(\frac{GM}{c^2}\right)^2
 \widehat\sigma_{\rm abs}\!\left[
 \frac{2\pi GMf_0(1+z)}{c^3}\right].
 \label{eq:populationconvolution}
\end{equation}
We use a flat $\Lambda$CDM benchmark with $H_0=67.4\,
\mathrm{km\,s^{-1}\,Mpc^{-1}}$, $\Omega_m=0.315$, and
$\Omega_{\rm DM}=0.264$~\cite{Planck2018VI}, and integrate to $z_{\max}=10$.
The response $\widehat\sigma_{\rm abs}$ is constructed directly from the
numerical transmission probabilities through $\ell_{\max}=12$.  Appendix
\ref{app:absorptionconvergence} shows that $\ell_{\max}=8$ already agrees with
this reference to better than $4\times10^{-10}$ over the full frequency grid.
To avoid extrapolating the calculation, it is set to zero outside the computed interval
$0.035\leq M\omega\leq1.2$.  This makes the tails conservative while leaving
the peaks inside the numerical domain.

\begin{figure}[t]
 \centering
 \includegraphics[width=0.88\linewidth]{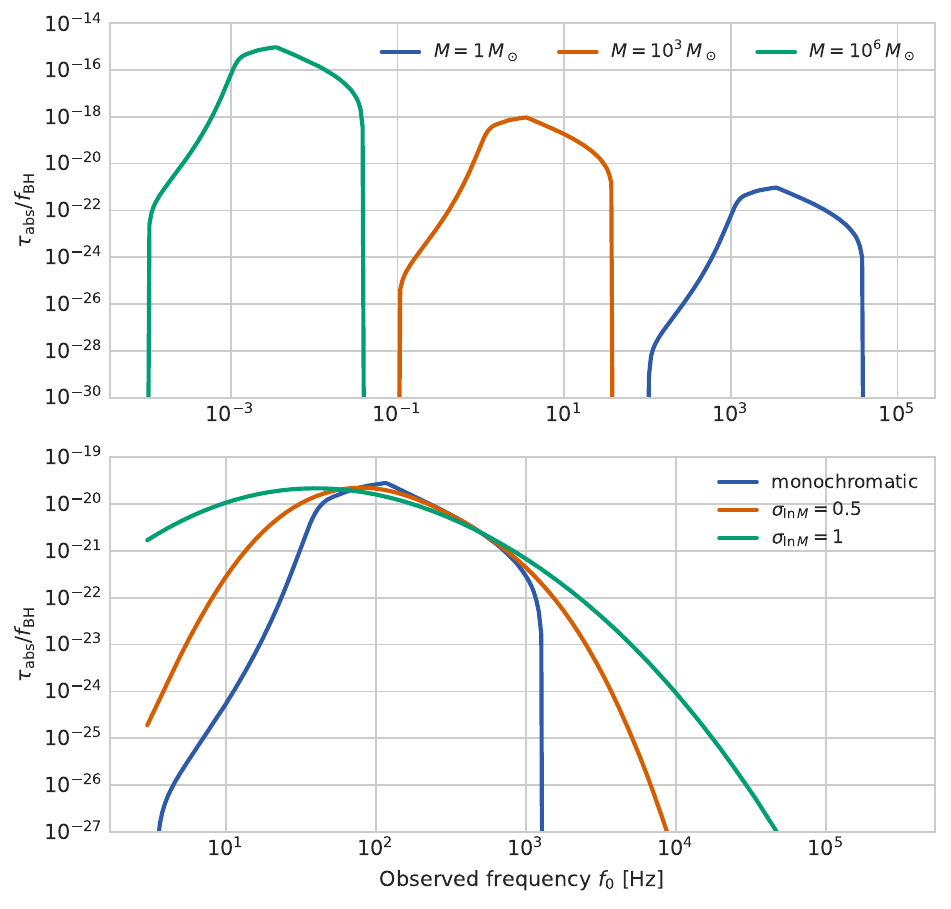}
 \caption{Cosmological absorption depth per unit black-hole dark-matter
 fraction, integrated to $z=10$.  Upper panel: monochromatic populations,
 illustrating the linear mass scaling and inverse shift of the characteristic
 frequency.  Lower panel: a $30M_\odot$ population with increasing lognormal
 width.  Redshift and mass dispersion broaden the single-hole response but do
 not overcome the very small optical depth.}
 \label{fig:cosmologicaldepth}
\end{figure}

For an extended population we adopt the illustrative lognormal mass-fraction
distribution
\begin{equation}
 \psi(M)=\frac{1}{\sqrt{2\pi}\sigma_{\ln M}}
 \exp\!\left[-\frac{\ln^2(M/M_c)}{2\sigma_{\ln M}^2}\right].
 \label{eq:lognormalmass}
\end{equation}
Figure~\ref{fig:cosmologicaldepth} shows both monochromatic populations and
lognormal examples.  For $f_{\rm BH}=1$, the peak depths are
$9.47\times10^{-22}$, $9.40\times10^{-19}$, and $9.34\times10^{-16}$ for
$M=1,10^3,10^6M_\odot$, respectively.  Their observed peak frequencies are
$3.52\,\mathrm{kHz}$, $3.45\,\mathrm{Hz}$, and $3.38\,\mathrm{mHz}$.  The
near-linear mass scaling and inverse frequency scaling follow directly from
Eq.~\eqref{eq:populationconvolution}.  They also provide a numerical check of
the dimensional argument in Eq.~\eqref{eq:alphascaling}.

For $M_c=30M_\odot$, broadening the distribution from monochromatic to
$\sigma_{\ln M}=0.5$ and $1$ shifts the peak from $116\,\mathrm{Hz}$ to
$79.2\,\mathrm{Hz}$ and $39.6\,\mathrm{Hz}$, while changing its height from
$2.83\times10^{-20}$ to $2.25\times10^{-20}$ and $2.19\times10^{-20}$.
The broad mass function consequently smears the spectral filter and moves
weight toward lower observed frequency.  The largest peak in this set,
$9.34\times10^{-16}$, is the quantitative population bound used below;
realistic abundance fractions multiply the curves downward.

\subsection{Population-specific normalizations}
\label{subsec:populationnormalizations}

The $f_{\rm BH}=1$ curves are useful scaling references but do not describe
known astrophysical populations.  We therefore show scenario envelopes rather
than precision curves.  Stellar remnants use a lognormal distribution with
$M_c=10M_\odot$ and $\sigma_{\ln M}=0.5$.  Their normalization is bracketed by
the legacy cosmic-inventory value $\Omega_{\rm sBH}=7\times10^{-5}$
\cite{FukugitaPeebles2004} and the modern stellar- and binary-evolution result
$\Omega_{\rm sBH}\simeq4\times10^{-4}$~\cite{Sicilia2022}.  For massive
galactic-nucleus black holes we adopt the illustrative local density
$\rho_{\rm MBH}=5.2\times10^5M_\odot\,\mathrm{Mpc}^{-3}$ and an illustrative
$M_c=10^8M_\odot$, $\sigma_{\ln M}=1$ distribution
\cite{GrahamDriver2007}.  Finally, as a constrained PBH example we use
$M_c=30M_\odot$, $\sigma_{\ln M}=0.5$, and $f_{\rm PBH}=10^{-3}$, representative
of the O3 merger-population upper limit across $1$--$200M_\odot$
\cite{AndresCarcasona2024}.

\begin{figure}[t]
 \centering
 \includegraphics[width=0.88\linewidth]{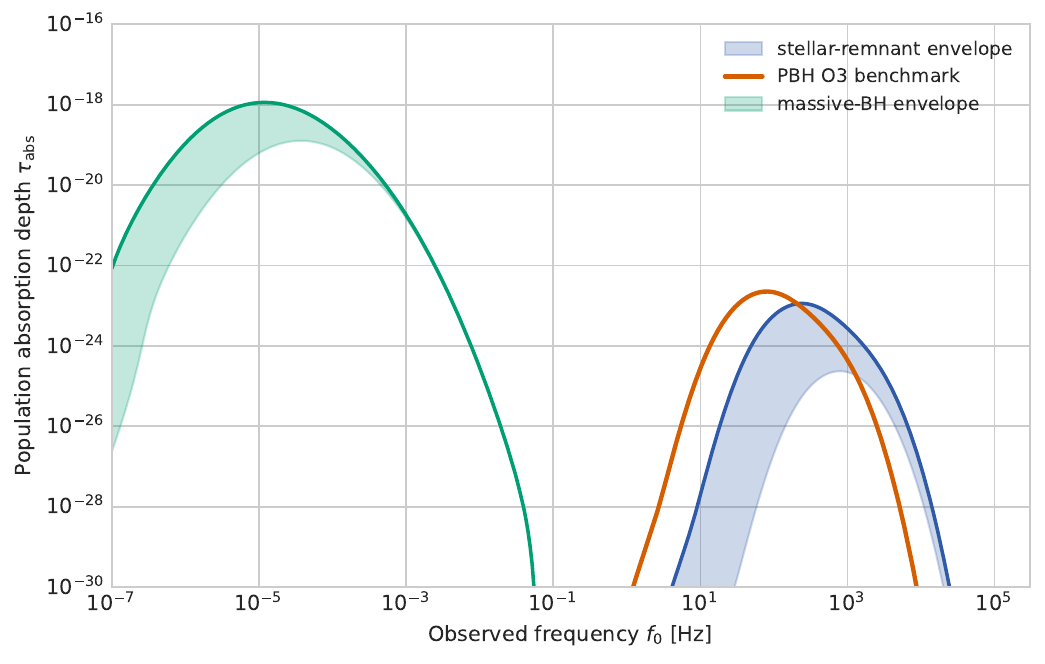}
 \caption{Absorption depths after population-specific normalization.  Shaded
 bands are scenario envelopes rather than statistical errors.  Their upper
 edges hold present-day astrophysical densities fixed to $z=10$; lower edges
 truncate them above $z=2$, with the stellar band additionally spanning the
 legacy-to-modern density normalization.  The PBH curve uses an illustrative
 O3 upper envelope and conserved comoving abundance.}
 \label{fig:specificpopulations}
\end{figure}

Figure~\ref{fig:specificpopulations} shows the resulting depths as sensitivity
envelopes, not confidence intervals.  The upper edges hold the present-day
comoving density fixed to $z=10$; the lower edges truncate the astrophysical
populations above $z=2$, and for stellar remnants also use the legacy lower
normalization.  The resulting peak ranges are
$2.34\times10^{-25}$--$1.14\times10^{-23}$ for stellar remnants and
$1.26\times10^{-19}$--$1.14\times10^{-18}$ for massive black holes.  Their
peak locations range from $778$ to $239\,\mathrm{Hz}$ and from $36.1$ to
$11.9\,\mu\mathrm{Hz}$, respectively.  The PBH benchmark, whose comoving
abundance is conserved, peaks at $2.25\times10^{-23}$ near $78.7\,\mathrm{Hz}$.
The massive population gives the largest depth because
Eq.~\eqref{eq:alphascaling} weights mass density by an additional factor of
$M$.  The sharp $z=2$ truncation is only a transparent measure of assembly-history
sensitivity; a fitted formation history would lie between model-dependent
alternatives.  Even the deliberately generous upper edges remain negligible.
The PBH normalization is also only a transparent
stellar-mass benchmark, not a universal constraint envelope: outside the LVK
mass interval, microlensing, evaporation, dynamical, and accretion bounds must
be applied separately.

\subsection{Asteroid-mass PBHs and Planck-scale relics}
\label{subsec:asteroidrelics}

The abundance and the absorption efficiency pull in opposite directions at
small mass.  PBHs in the approximate interval
$10^{-16}$--$10^{-12}M_\odot$ may still constitute all of the dark matter
\cite{CarrKohri2021,Ballesteros2025PBH}.  Setting $f_{\rm PBH}=1$, we find the
peak values summarized below (Table~\ref{tab:asteroidpbhpeaks}).
\begin{table}[ht]
\centering
\caption{Peak observed absorption frequency and cosmological absorption depth
for monochromatic asteroid-mass PBH populations with $f_{\rm PBH}=1$.}
\label{tab:asteroidpbhpeaks}
\begin{tabular}{ccc}
 \toprule
 $M/M_\odot$ & peak $f_0$ [Hz] & peak $\tau_{\rm abs}$\\
 \midrule
 $10^{-16}$ & $3.47\times10^{19}$ & $9.43\times10^{-38}$\\
 $10^{-14}$ & $3.47\times10^{17}$ & $9.43\times10^{-36}$\\
 $10^{-12}$ & $3.47\times10^{15}$ & $9.43\times10^{-34}$\\
 \bottomrule
\end{tabular}
\end{table}
Thus the open abundance window does not produce a large opacity.  At fixed
dark-matter density, the number density grows as $M^{-1}$ but the absorption
cross section falls as $M^2$, leaving $\tau_{\rm abs}\propto M$.  Moreover,
the characteristic absorption frequency scales as $M^{-1}$ and lies in the
petahertz-to-exahertz domain.  This should not be confused with the much lower
frequency of the induced stochastic background associated with the formation
of asteroid-mass PBHs.

Stable relics left when Hawking evaporation terminates near the Planck scale
have also been proposed as all of the dark matter
\cite{CarrKohri2021,LehmannProfumo2019}.  Substituting
$M=M_{\rm Pl}=2.18\times10^{-8}\,\mathrm{kg}=1.09\times10^{-38}M_\odot$ into
the classical scaling gives a \emph{formal} peak
\begin{equation}
 f_{0,\rm peak}\simeq3.13\times10^{41}\,\mathrm{Hz},\qquad
 \tau_{\rm abs}\simeq1.03\times10^{-59}
 \quad(f_{\rm relic}=1).
 \label{eq:relicformal}
\end{equation}
These numbers are not a semiclassical prediction.  A Planck-mass object has a
Schwarzschild radius of order the Planck length, and both its internal structure
and its interaction with Planck-frequency gravitational disturbances require
quantum gravity.  Equation~\eqref{eq:relicformal} is included only to show how
violently the classical opacity would be suppressed if the $M^2$ cross-section
scaling persisted.  The relics may behave more like particles than classical
black holes, so model-dependent non-geometric interactions are outside the
scope of this transfer calculation.

\begin{figure}[t]
 \centering
 \includegraphics[width=0.88\linewidth]{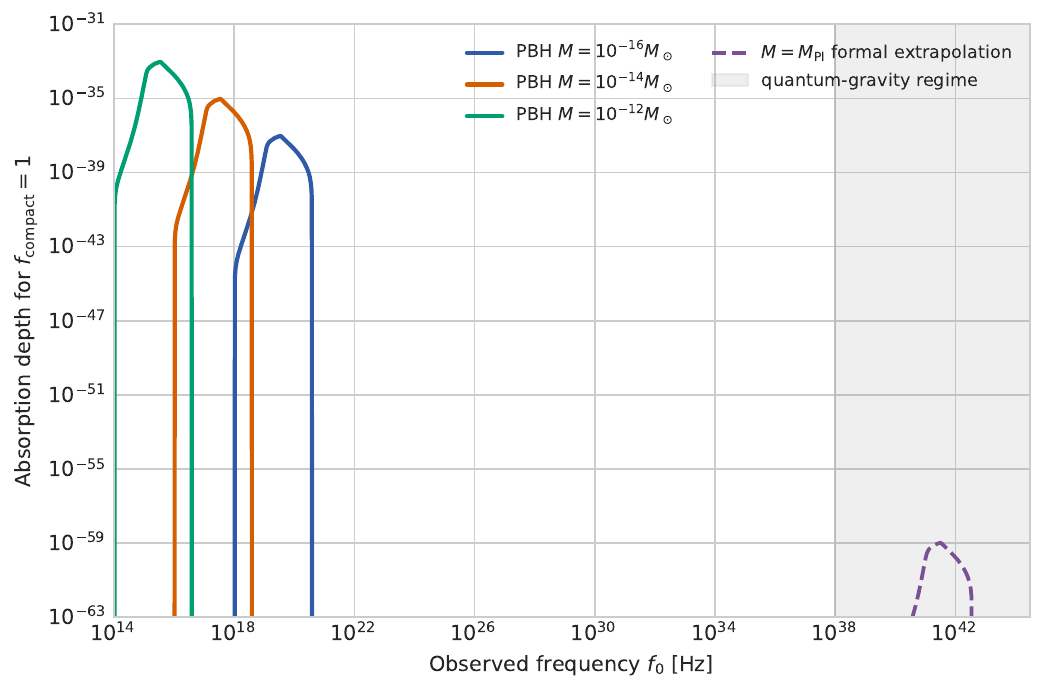}
 \caption{Absorption depth for PBHs comprising all dark matter in the
 asteroid-mass window, together with the formal continuation to a Planck-mass
 relic.  The shaded high-frequency region emphasizes that the relic curve lies
 outside the domain of semiclassical black-hole perturbation theory.}
 \label{fig:asteroidrelics}
\end{figure}

\section{Discussion}
\label{sec:discussion}

The calculations give a clean three-way distinction.  A stationary
Schwarzschild hole is a passive filter: its absorption probability has a broad
quadrupolar turnover and its complex reflection phase has a sharper QNM-scale
delay, but it creates no independent spectral line.  Finite illumination
generates delayed ringdown whose energy was supplied by the incident packet.
A rotating hole adds a third possibility: a co-rotating circular mode in the
superradiant band receives stationary gain powered by the hole's spin.

The novelty is consequently in the synthesis and in a limited set of new
calculations, not in the existence of black-hole greybody factors or spin-$4$
SGWB polarization.  Relative to the low-energy transport treatment of
Ref.~\cite{Pizzuti2023}, we supply the parity-resolved complex kernel through
the QNM-scale regime, include horizon absorption, and perform an explicit
redshift--mass-function opacity convolution.  Relative to strong-field
waveform scattering~\cite{LiHouZhao2025}, we formulate a stochastic collision
operator and its population scaling.  The reflected-channel Wigner-delay curve
provides a phase-sensitive diagnostic complementary to existing QNM--greybody
relations.  Its center is robust under the numerical and fit variations tested,
whereas its descriptive width is model-dependent at the 10--20\% level.  The
separate AAA continuation resolves that ambiguity by extracting both pole
components to sub-percent median accuracy; it remains a cross-check of the same
scattering solution, not an independent perturbation code or a forward-axis
solution.

At population level, all three effects inherit the same geometric scarcity.
The optical depth rules out a large cumulative attenuation for the homogeneous
Schwarzschild benchmarks, including asteroid-mass PBHs comprising all dark
matter.  Superradiance changes the sign and can change the modal coefficient
by order unity, but it does not change the decisive $\rho_{\rm BH}ML$ scaling.
Its gain is also chiral and aligned rather than isotropic.  The on-axis
estimate in Eq.~\eqref{eq:kerrtau}, which omits this ensemble dilution, is
already negligible.

The elastic angular result carries an analysis-dependent forward response, but
not an uncontrolled order-of-magnitude conclusion.  Replacing the illustrative
$20^\circ$ hard cut by cutoffs from $1^\circ$ to $30^\circ$ changes the
low-frequency transport moment by factors between 4.28 and 0.63
(Table~\ref{tab:cutoffsensitivity}); the corresponding population depth remains
below $3\times10^{-22}f_{\rm BH}(M/M_\odot)$.  More importantly, the isotropic
monopole result used for the absorption curves is exactly independent of this
choice.

The negative population result is part of the conclusion, not a failure of
the single-hole calculation.  The most credible observational targets are
rare strong-field configurations: transient waves passing close to a hole, or
directionally and helically selected waves interacting with a rapidly spinning
hole.  The explicitly schematic encounter in
Appendix~\ref{app:rareencounter} illustrates the required scale, but its
isotropic redistribution assumption is not a sourced Teukolsky prediction and
is not used as a quantitative constraint.
Such a signal can be either enhanced or suppressed by interference.  It is not
a large broadband distortion accumulated through a homogeneous cosmological
population.

\section{Conclusions}

Stationary GW excitation of a Schwarzschild black hole is consistently
described by a real-frequency scattering transfer function.  QNM poles shape
that function but do not supply an additional stationary emission line.  The
amplitude has a broad absorption transition near the quadrupolar QNM scale and
a sharper phase signature, whose Wigner delay peaks at
$M\omega=0.3755$ with $\tau_2/M=14.54$ in the fiducial analysis.  The peak
location is numerically robust, while a pole-profile width is only a descriptive
QNM-scale diagnostic.  Direct AAA continuation of the complex reflection
amplitude instead yields $M\omega_{20}=0.37389-0.08932i$, in sub-percent median
agreement with the reference pole across the stability audit.  Its finite-angle Mueller kernel
redistributes anisotropy and polarization while conserving the elastic
intensity monopole; in particular, unpolarized anisotropy sources $E$ but not
$B$ polarization for a Schwarzschild population.

A causal calculation supplies the complementary emission statement.  Finite
incident packets centered below, near, and above the barrier generate delayed
outgoing waveforms whose fitted complex frequencies agree with the fundamental
quadrupolar QNM.  For the packet centered at $M\omega_0=0.37367$, about $1\%$
of the incident energy crosses the extraction point after the conservative
late-time start used here.  This is transient re-emission of incident energy,
not a stationary line or amplification.

The same framework fixes the population scaling.  The single-hole far-field
power is proportional to $M^2/r^2$, and an incoherent population is governed by
the optical depth rather than a nearest-neighbour cutoff.  The absorption depth
is proportional to $\rho_{\rm BH}ML$ and is extremely small for the Galactic
and cosmological benchmarks studied here.  Allowing the stellar-remnant
normalization and a transparent $z=2$--10 assembly bracket produces peak
depths of order $10^{-25}$--$10^{-23}$, while the massive-black-hole bracket
is of order $10^{-19}$--$10^{-18}$.  These are scenario ranges rather than
statistical intervals.  This remains true for
asteroid-mass PBHs comprising all dark matter; Planck relics cannot be assessed
within semiclassical perturbation theory.

Kerr rotation supplies genuine stationary amplification in the co-rotating
circular channel, reaching the known modal maximum $Z_{222}\simeq1.38$ near
extremality.  That maximum is not an isotropic-background gain: opposite
helicity, incidence angle, and random spin orientation dilute it.  Even the
aligned numerical benchmark gives only
$\tau_{\rm gain}^{\rm axis}\simeq2.0\times10^{-24}f_{\rm BH}(M/M_\odot)$ over
a cosmic-mean Hubble path.

The final result is therefore sharp.  Black-hole response to incident GWs is
rich and physically distinct in its passive, transient, and superradiant
channels, but a homogeneous population produces no appreciable SGWB
distortion in the regimes studied here.  The transfer kernel, causal ringdown,
and Kerr mode selection are the robust single-hole predictions.  Any viable
observational application should target rare, nearby, aligned, rapidly
spinning, or transiently illuminated systems rather than cumulative
cosmological propagation.  An order-ten-percent strain effect appears only in
the deliberately schematic strong-field construction of
Appendix~\ref{app:rareencounter}.

\section*{Data Availability Statement}

The numerical data and source code required to reproduce all figures and
quoted numerical results are contained in the anonymous ancillary archive
supplied with this submission under the filename
\nolinkurl{Black_Hole_Excitations_anonymous_review.zip}.  It includes machine-readable tables, figure-generation
scripts, and a reproducibility manifest.  The same version will be deposited
in a permanent public repository upon acceptance.  No proprietary or
restricted data were used.

\acknowledgments

SP thanks Stefano Liberati for asking the question that spurred this line of investigation. 
This work was supported by the U.S. Department of Energy, Office of Science,
Office of High Energy Physics, under Award Number DE-SC0010107, and by a
ChatGPT for Academic Researchers grant from OpenAI.  OpenAI ChatGPT and Codex,
accessed during manuscript preparation, were used as research-assistance tools for code
development, numerical auditing, workflow documentation, and editorial
revision.  The authors critically reviewed and validated all AI-assisted
material and remain fully responsible for the accuracy, integrity,
originality, and scientific content of the manuscript.

\appendix
\section{Schematic strong-field Kerr encounter}
\label{app:rareencounter}

The homogeneous optical depth can be negligible even when a single contrived
geometry produces a visible response.  Consider an intentionally favorable
encounter: a pure co-rotating circular $(\ell,m)=(2,2)$ wave is incident along
the spin of a nearly extremal Kerr hole.  Using the computed on-axis benchmark
$\sigma_{\rm gain}^{\rm axis}=6M^2$, and making the crude assumption that the
extracted power is redistributed isotropically, inverse-square propagation
would give
\begin{equation}
 \frac{h_{\rm gain}}{h_{\rm dir}}
 \simeq \frac{1}{R}\sqrt{\frac{\sigma_{\rm gain}^{\rm axis}}{4\pi}}
 =\frac{0.69M}{R}.
 \label{eq:rarestrain}
\end{equation}
This is a geometric proxy, not a physical emission pattern.  The true
scattered field inherits the spin-weighted $(2,2)$ angular structure and a
complex, direction-dependent Teukolsky amplitude.  Depending on observer
angle, replacing isotropic redistribution by that pattern can enhance or
suppress the inferred strain and removes any universal interpretation of a
single separation threshold.

\begin{figure}[ht]
 \centering
 \includegraphics[width=0.96\linewidth]{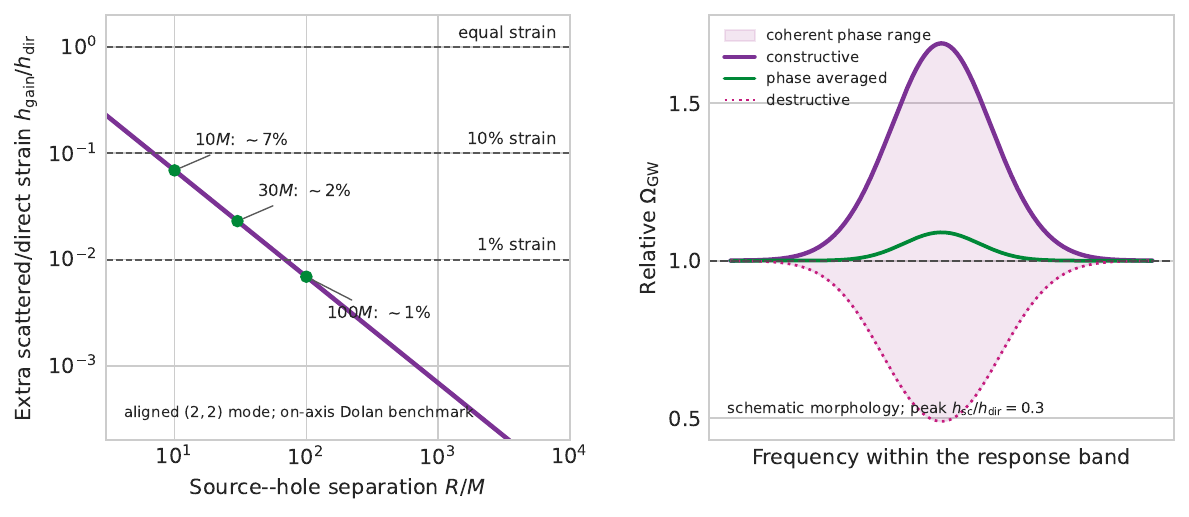}
 \caption{Schematic strong-field Kerr encounter.  Left: geometric strain
 scaling under the crude isotropic-redistribution assumption.  Right:
 uncalibrated spectral topologies for a localized component with peak strain
 ratio $0.3$.  Neither panel is a sourced Teukolsky waveform or angular
 transfer calculation, and the figure is not used in the population bounds.}
 \label{fig:rareencounter}
\end{figure}

The left panel of Fig.~\ref{fig:rareencounter} displays the proxy scaling.
Its nominal values are $6.9\%$, $2.3\%$, and $0.7\%$ at
$R/M=10,30,$ and $100$; these label the assumed geometry rather than robust
predictions.  If $q(f)$ denotes the unspecified scattered-to-direct strain
transfer and $\phi(f)$ its relative phase, then
\begin{equation}
 \OmGW^{\rm obs}(f)=\OmGW^{(0)}(f)
 \left|1+q(f)e^{i\phi(f)}\right|^2.
 \label{eq:illustrativetransfer}
\end{equation}
For the schematic peak $q=0.30$, constructive interference, phase averaging,
and opposite coherent phase give ratios $1.69$, $1.09$, and $0.49$,
respectively.  A quantitative prediction requires a sourced Teukolsky
calculation retaining the direct--scattered cross spectrum.

\clearpage
\section{Angular-kernel diagnostics}
\label{app:diagnostics}

\begin{figure}[ht]
 \centering
 \includegraphics[width=0.72\linewidth]{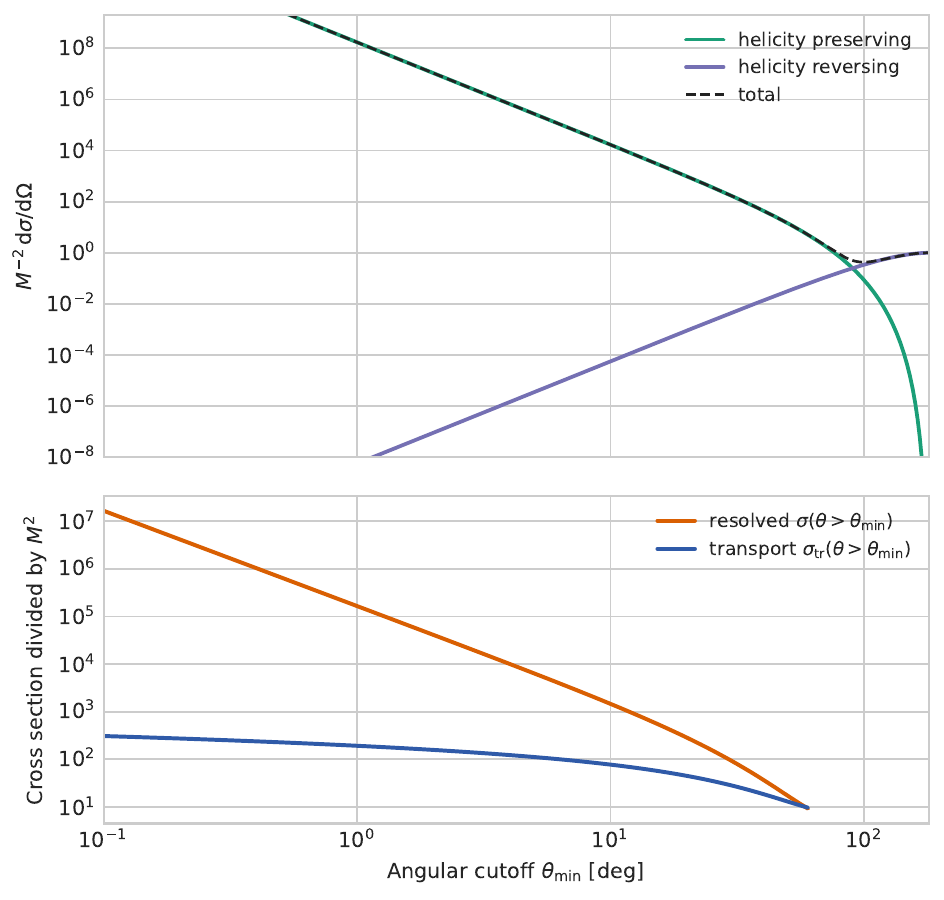}
 \caption{Low-frequency helicity-preserving and helicity-reversing kernels and
 their resolved and transport-weighted moments.  The curves expose the cutoff
 dependence of the coherent forward contribution.}
 \label{fig:angularkernel}
\end{figure}

This appendix collects visual diagnostics removed from the main narrative.
Figure~\ref{fig:angularkernel} displays the analytic low-frequency limit and
its cutoff dependence.  This is a validation regime, not a substitute for the
finite-frequency kernel used in the main text.

\begin{figure}[p]
 \centering
 \includegraphics[width=0.62\linewidth]{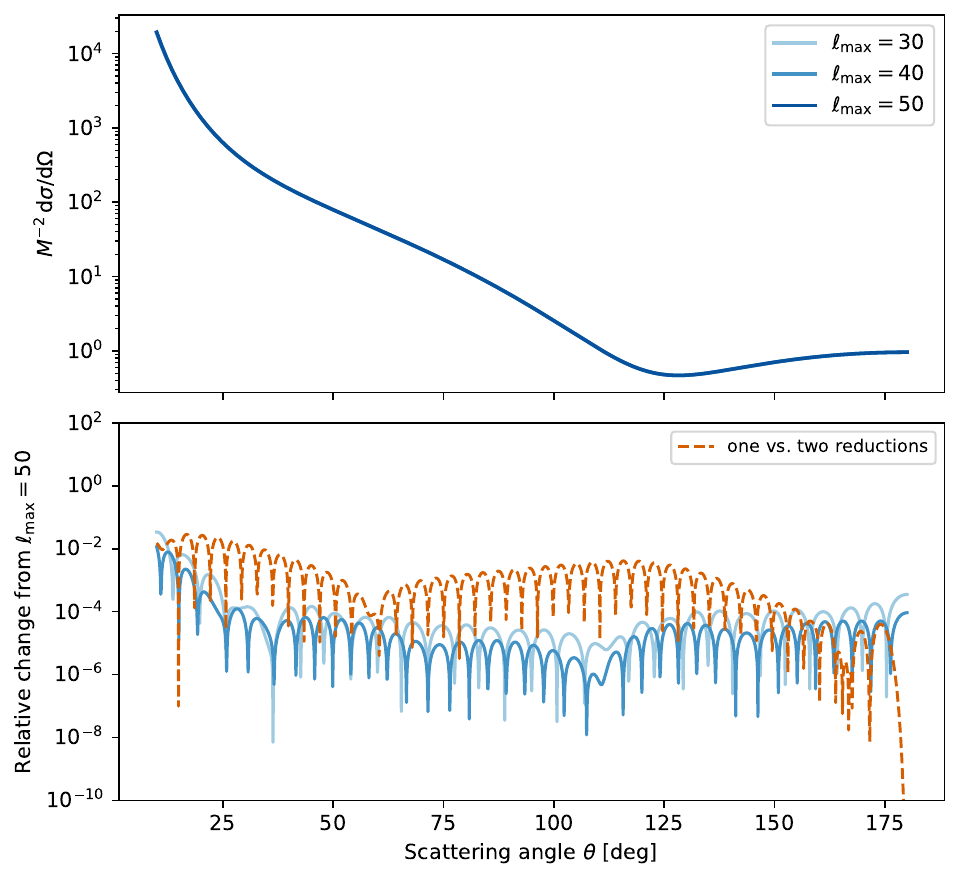}
 \caption{Finite-frequency angular-kernel convergence at
 $M\omega=0.37367$ after two series reductions.  Phase shifts through
 $\ell=60$ pad the displayed truncations through $\ell_{\max}=50$.}
 \label{fig:finitekernel}
 \vspace{0.4em}
 \includegraphics[width=0.62\linewidth]{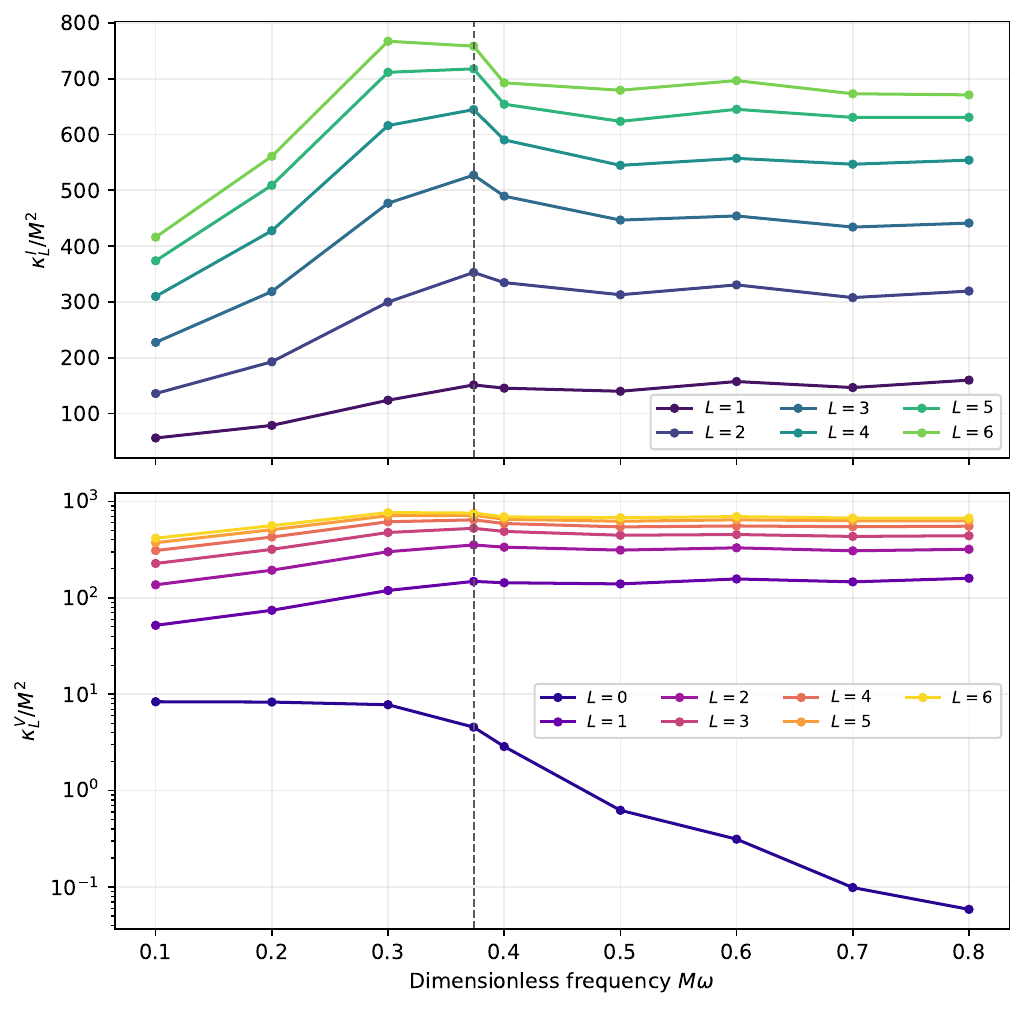}
 \caption{Resolved intensity and circular-polarization collision eigenvalues
 for $\theta_{\min}=20^\circ$.  The identities
 $\kappa_0^I=0$, $\kappa_1^I=\sigma_{\rm tr,>}$, and
 $\kappa_0^V=2\sigma_{g,>}$ provide normalization checks.}
 \label{fig:boltzmannmultipoles}
\end{figure}

At $M\omega=0.37367$, increasing the physical partial-wave cutoff from
$\ell_{\max}=40$ to 50 changes $M^{-2}\dd\sigma/\dd\Omega$ by less than
$3.4\times10^{-4}$ for $\theta\ge20^\circ$.  One versus two series reductions
changes the result by $2.5\%$ at $20^\circ$, below $1\%$ at $30^\circ$, and
below $0.4\%$ at $45^\circ$.  Figure~\ref{fig:finitekernel} shows the full
angular comparison.

\begin{figure}[p]
 \centering
 \includegraphics[width=0.72\linewidth]{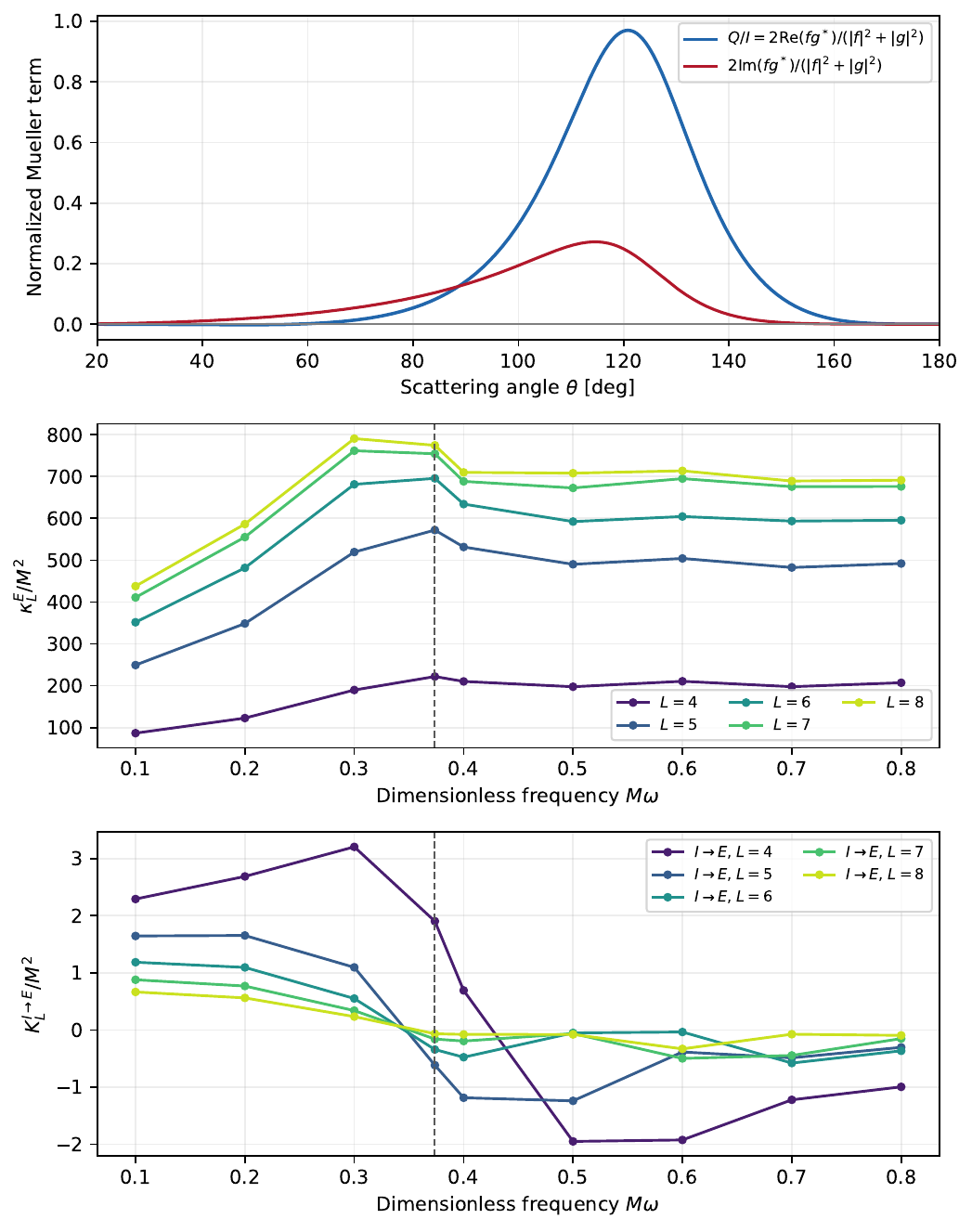}
 \caption{Phase-aware polarization transfer above $20^\circ$: local Mueller
 terms, spin-4 $E$-mode damping, and the scalar-intensity to $E$-mode source.
 Schwarzschild parity forbids an $I\to B$ source.}
 \label{fig:muellerspin4}
\end{figure}

\begin{figure}[p]
 \centering
 \includegraphics[width=0.76\linewidth]{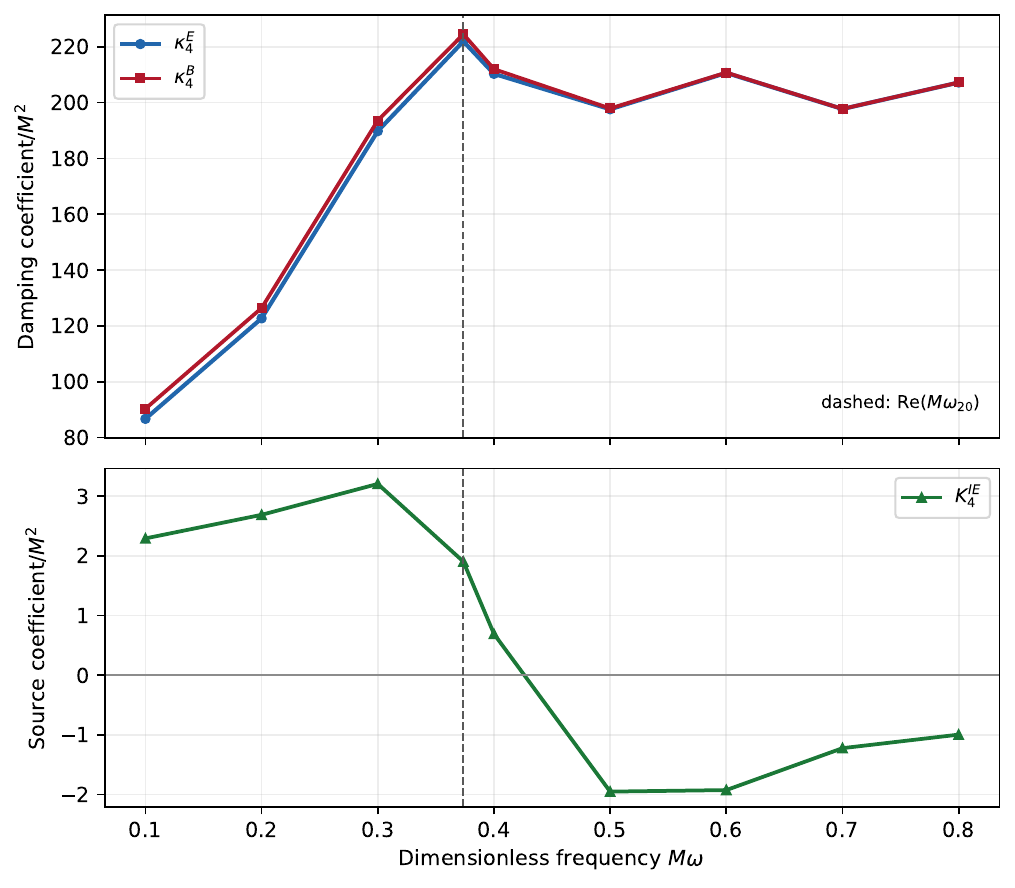}
 \caption{Frequency dependence of the principal $L=4$ polarization-transfer
 coefficients for $\theta_{\min}=20^\circ$.  The dashed line marks the real
 part of the fundamental Schwarzschild quadrupolar QNM frequency.}
 \label{fig:spin4frequency}
\end{figure}

Figure~\ref{fig:spin4frequency} isolates the principal quadrupolar
polarization-transfer quantities over $0.1\leq M\omega\leq0.8$ at the same
$20^\circ$ cutoff.  The two damping eigenvalues track one another closely but
are not constant: they rise toward the QNM-scale region and retain
finite-frequency oscillations above it.  The scalar $I\to E$ source is much
smaller and changes sign between $M\omega=0.4$ and $0.5$.  Thus the values in
Eq.~\eqref{eq:spin4numbers} are a benchmark slice through a frequency-dependent
kernel, not universal constants.

\clearpage
\section{Conventions and normalization audit}
\label{app:normalizationaudit}

For reproducibility, Table~\ref{tab:normalizationaudit} collects the closure
tests used to audit the factors and signs in the scattering-to-transport map.
We define the circular-basis coherency matrix by
$I=H_{++}+H_{--}$, $V=H_{++}-H_{--}$, and
$Q-iU=2H_{+-}$.  Acting on this matrix with
$J=\left(\begin{smallmatrix}f&g\\g&f\end{smallmatrix}\right)$ reproduces
Eq.~\eqref{eq:muellermatrix}, including the signs in its $U$--$V$ block.
With this choice $Q+iU$ has spin weight $-4$; reversing the handedness
convention changes the signs of $U$, $V$, and the named parity-odd coefficient
together, but leaves intensities and measurable polarization invariant.

The factor $2\pi$ in the collision projections is the azimuthal integral for
an isotropic population, not an additional polarization average.  The incident
plane-wave normalization is already contained in $f$ and $g$, for which
$\dd\sigma/\dd\Omega=|f|^2+|g|^2$.  Likewise,
Eq.~\eqref{eq:absorptioncrosssection} includes both parity sectors: their equal
Schwarzschild transmission probabilities turn the conventional
$\pi/(2\omega^2)$ parity sum into the displayed $\pi/\omega^2$ expression.

\begin{table}[ht]
 \centering
 \caption{Independent normalization and convention checks.  ``Machine''
 denotes double-precision closure; the dipole comparison is limited by the
 independently sampled angular integrations.}
 \label{tab:normalizationaudit}
 \begin{tabular}{p{0.47\linewidth}p{0.35\linewidth}}
  \toprule
  Check & Result \\
  \midrule
  Direct coherency propagation versus Mueller matrix & machine precision \\
  $S^2-A^2=D^2+B^2$ point by point & $<8.9\times10^{-16}$ \\
  $d^4_{44}$, $d^4_{4,-4}$, $d^4_{40}$ versus closed forms & machine precision \\
  Elastic intensity monopole & $\kappa_0^I=0$ \\
  Helicity-reversal monopole & $\kappa_0^V=2\sigma_{g,>}$ \\
  Intensity dipole versus transport moment & $\kappa_1^I=\sigma_{\rm tr,>}$ \\
  Dimensional scaling & $f,g\propto M$ and $\sigma,\kappa\propto M^2$ \\
  \bottomrule
 \end{tabular}
\end{table}

\section{Numerical flux check}

\begin{figure}[ht]
 \centering
 \includegraphics[width=0.70\linewidth]{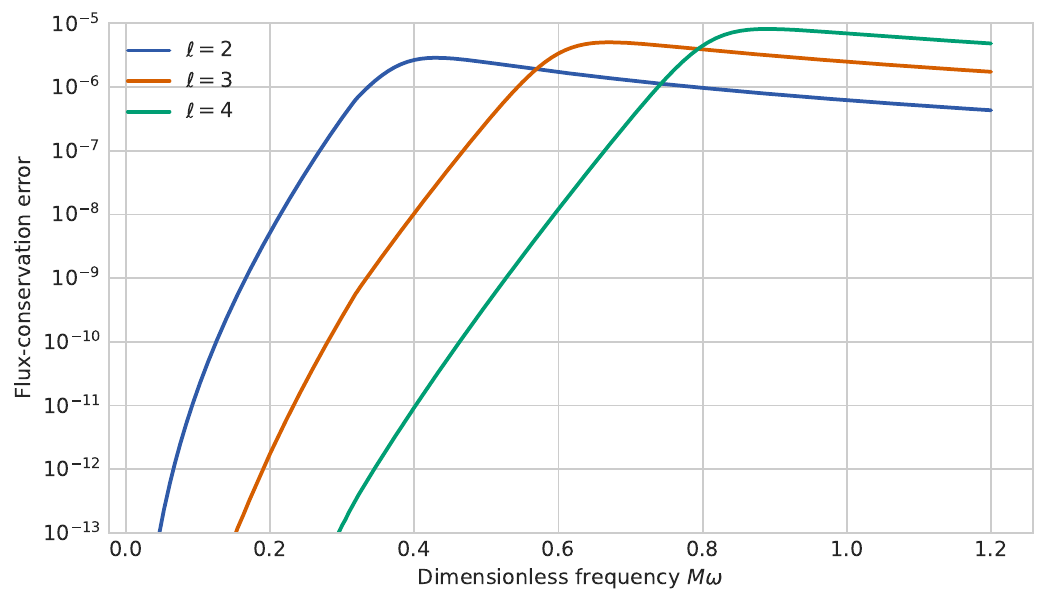}
 \caption{Absolute flux-conservation error across the numerical grid.}
 \label{fig:convergence}
\end{figure}

Figure~\ref{fig:convergence} displays the absolute defect in
Eq.~\eqref{eq:flux}.  Its smooth frequency dependence reflects the finite
starting and extraction radii.  Increasing both domains reduces the defect.

\subsection{Absorption multipole convergence}
\label{app:absorptionconvergence}

\begin{figure}[ht]
 \centering
 \includegraphics[width=0.76\linewidth]{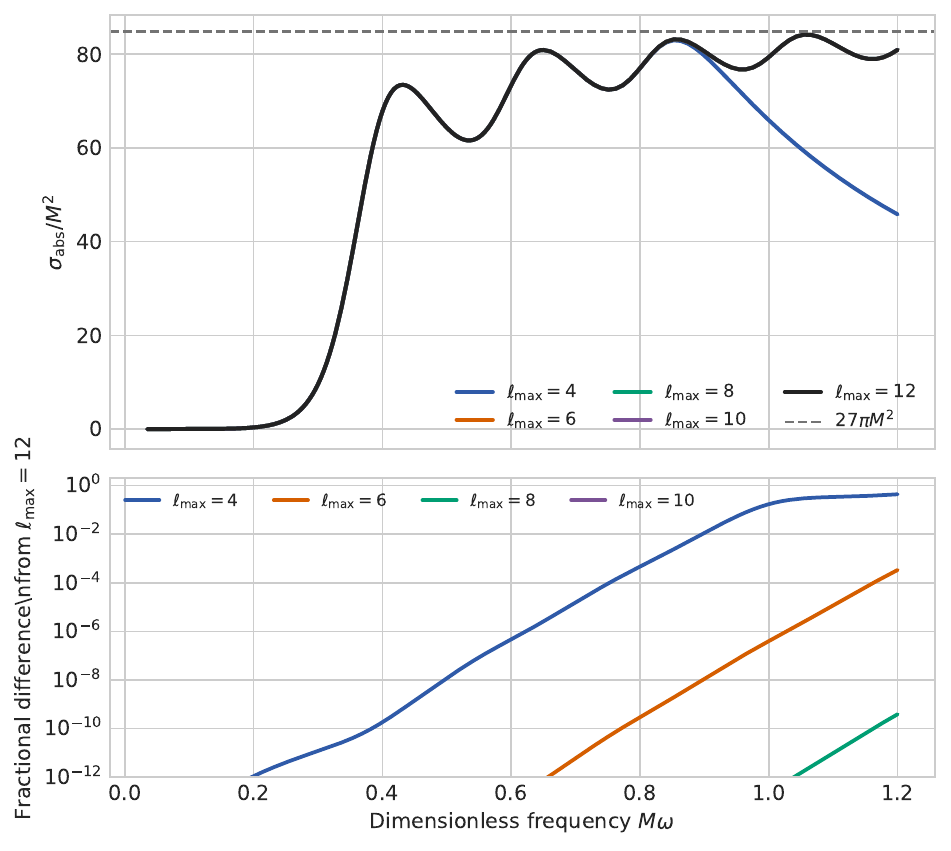}
 \caption{Convergence of the Schwarzschild absorption cross section with the
 maximum included multipole.  Top: cumulative partial-wave sums and the
 geometric-optics limit.  Bottom: fractional difference from the
 $\ell_{\max}=12$ reference.}
 \label{fig:absorptionconvergence}
\end{figure}

The population calculation requires the partial-wave sum in
Eq.~\eqref{eq:absorptioncrosssection}, rather than only the leading modes shown
in Fig.~\ref{fig:response}.  Figure~\ref{fig:absorptionconvergence} compares
cumulative sums through $\ell_{\max}=4,6,8,10,$ and $12$.  Relative to the
$\ell_{\max}=12$ reference, their maximum differences over
$0.035\leq M\omega\leq1.2$ are $0.434$, $3.35\times10^{-4}$,
$3.91\times10^{-10}$, and $1.8\times10^{-16}$, respectively.  Thus the former
$\ell\leq4$ truncation was inadequate only toward the upper end of the grid;
$\ell_{\max}=8$ is already amply converged throughout it.  At $M\omega=1.2$
the converged cross section is $80.95M^2$, only $4.57\%$ below the
geometric-optics limit $27\pi M^2$; the remaining offset is the expected
finite-frequency oscillation, not a missing high-$\ell$ tail.

\subsection{Time-domain convergence}
\label{app:tdconvergence}

\begin{figure}[ht]
 \centering
 \includegraphics[width=0.76\linewidth]{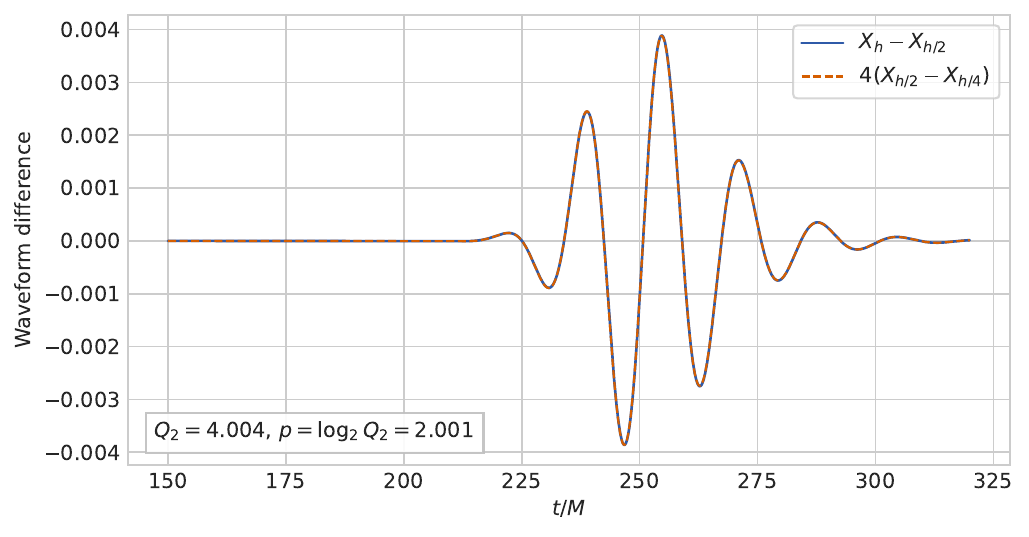}
 \caption{Second-order convergence of the extracted time-domain waveform.
 The difference between the two coarser solutions is compared with four times
 the difference between the two finer solutions, at fixed Courant ratio.}
 \label{fig:tdconvergence}
\end{figure}

We test the leapfrog evolution on three nested grids,
$\Delta r_*/M=(0.20,0.10,0.05)$, holding
$\Delta t/\Delta r_*=0.5$ fixed and otherwise using the central packet of
Sec.~\ref{subsec:wavepacket}.  After restriction to the coarsest time grid, we
define $D_{h,h/2}$ as the root-mean-square waveform difference over
$150\leq t/M\leq320$.  The measured norms are
$D_{h,h/2}=1.06706\times10^{-3}$ and
$D_{h/2,h/4}=2.66491\times10^{-4}$, hence
\begin{equation}
 Q_2=\frac{D_{h,h/2}}{D_{h/2,h/4}}=4.004,
 \qquad p=\log_2 Q_2=2.001.
\end{equation}
Figure~\ref{fig:tdconvergence} also shows the expected pointwise collapse of
$X_h-X_{h/2}$ onto $4(X_{h/2}-X_{h/4})$.  The corresponding fitted
$(M\omega,M\gamma)$ values are $(0.371722,0.088406)$,
$(0.371706,0.088409)$, and $(0.371704,0.088417)$; their small non-monotonic
scatter is subdominant to the fit-window dependence and is not used to infer a
separate convergence order.

\begin{table}[ht]
 \centering
 \caption{Consolidated numerical-stability and sensitivity budget.  Entries
 are changes under the stated variation, not statistical confidence
 intervals.  The delay-profile background and series-reduction comparison are
 method sensitivities; the stellar-remnant row is an astrophysical
 normalization systematic; the remaining entries are numerical convergence or
 consistency tests.}
 \label{tab:uncertaintybudget}
 \begin{tabular}{@{}>{\raggedright\arraybackslash}p{0.25\linewidth}
 >{\raggedright\arraybackslash}p{0.34\linewidth}
 >{\raggedright\arraybackslash}p{0.31\linewidth}@{}}
  \toprule
  Quantity & Variation & Observed change or defect \\
  \midrule
  RW reflection/transmission & Full $\ell=2$--12, $M\omega=0.035$--1.2 grid
    & Maximum flux defect $1.3\times10^{-5}$ \\
  Absorption multipole sum & $\ell_{\max}=4,6,8,10,12$
    & Maximum difference from $\ell_{\max}=12$: $3.9\times10^{-10}$ at
      $\ell_{\max}=8$ \\
  Wigner-delay extraction radius & Increase $r_{\max}$ by $50\%$
    & $|\Delta\tau|<0.067M$ where $|\cR_2|^2>0.03$; peak bin unchanged \\
  Wigner-delay domain/tolerance & Double extraction radius; tighten ODE
    tolerance and move horizon start & Peak shift at most one $0.0015$ bin;
    height changes $0.08M$ and $5\times10^{-5}M$, respectively \\
  Phase smoothing & Savitzky--Golay windows 9--31 and orders 2--5
    & Peak fixed at $M\omega=0.3755$; height $14.39$--$14.55M$ \\
  Descriptive delay fit & Change fit interval and linear/quadratic background
    & $x_c=0.3718$--$0.3736$; $\Gamma=0.086$--$0.107$ \\
  AAA pole & 60 window, sampling, and tolerance combinations
    & At most $0.19\%$ in frequency and $1.29\%$ in damping rate \\
  Time-domain waveform & $\Delta r_*/M=(0.20,0.10,0.05)$ at CFL $=0.5$
    & $Q_2=4.004$; observed order $p=2.001$ \\
  Time-domain fitted pole & Same three nested grids
    & $M\omega=0.371704$--$0.371722$;
      $M\gamma=0.088406$--$0.088417$ \\
  Ringdown fit window & Start $2.5\sigma_p$--$3.5\sigma_p$; duration $40M$--$70M$
    & $M\omega=0.3711$--0.3729; $M\gamma=0.0880$--0.0892 \\
  Incident-packet width & $\sigma_p/M=6,8,10$ at
    $M\omega_0=0.37367$ & $M\omega=0.37012$--0.37239;
    $M\gamma=0.08836$--0.08850 \\
  Finite-angle phase shifts & Asymptotic order 10--13 at $\ell=60$
    & Phase changes $2.3\times10^{-9}$ rad; flux defect $<2.3\times10^{-10}$ \\
  Angular partial-wave sum & $\ell_{\max}=40$--50; one versus two reductions
    & Cutoff change $<3.4\times10^{-4}$ above $20^\circ$; reduction change
    $2.5\%$ at $20^\circ$, $<1\%$ at $30^\circ$ \\
  Population convolution & $N_z=401$--1601 and $N_{\ln M}=81$--161
    & Representative $30M_\odot$, $\sigma_{\ln M}=1$ peak changes $<0.1\%$ \\
  Stellar-remnant normalization & Legacy $\Omega_{\rm sBH}=7\times10^{-5}$
    versus modern population estimate $\simeq4\times10^{-4}$
    & Optical depth rescales linearly; factor-of-few astrophysical systematic \\
  Wigner-$d$/Mueller projection & Closed forms and coherency-matrix audit
    & Agreement at double-precision roundoff \\
  \bottomrule
 \end{tabular}
\end{table}

\clearpage

\section{Derivation of the cosmological optical depth}
\label{app:opticaldepthderivation}

For completeness, we derive Eq.~\eqref{eq:populationconvolution} from the
local attenuation law.  In the rest frame of a dilute absorber population, a
ray crossing a proper distance $\dd s=c\,|\dd t|$ has survival probability
\begin{equation}
 \frac{\dd I_f}{I_f}=-\dd\tau_f,
 \qquad
 \dd\tau_f=\dd s\int\dd\ln M\,
 \frac{\dd n_{\rm phys}}{\dd\ln M}(M,z)\,
 \sigma_{\rm abs}(f,M).
 \label{eq:localattenuation}
\end{equation}
This relation uses physical number density and the cross section measured in
the same local orthonormal frame.  In an FLRW spacetime,
$\dd t/\dd z=-[(1+z)H(z)]^{-1}$, while a wave observed at $f_0$ has local
frequency $f=(1+z)f_0$.  Hence
\begin{equation}
 \tau_{\rm abs}(f_0)=\int_0^{z_{\max}}
 \frac{c\,\dd z}{(1+z)H(z)}
 \int\dd\ln M\,
 \frac{\dd n_{\rm phys}}{\dd\ln M}(M,z)
 \sigma_{\rm abs}[(1+z)f_0,M].
 \label{eq:taucosmogeneral}
\end{equation}

The benchmark curves assume a conserved comoving population whose present
mass density is a fraction $f_{\rm BH}$ of the dark-matter density.  If
$\psi(M)$ is the mass fraction per $\dd\ln M$, normalized to unity, then
\begin{equation}
 \frac{\dd n_{\rm phys}}{\dd\ln M}
 =(1+z)^3\frac{\dd n_{\rm com}}{\dd\ln M}
 =f_{\rm BH}\rho_{\rm DM,0}(1+z)^3\frac{\psi(M)}{M}.
 \label{eq:physicalnumberdensity}
\end{equation}
Finally, Schwarzschild scale invariance gives
\begin{equation}
 \sigma_{\rm abs}(f,M)=\left(\frac{GM}{c^2}\right)^2
 \widehat\sigma_{\rm abs}\!\left(\frac{2\pi GMf}{c^3}\right).
 \label{eq:dimensionfulcrosssection}
\end{equation}
Substitution of Eqs.~\eqref{eq:physicalnumberdensity} and
\eqref{eq:dimensionfulcrosssection} into Eq.~\eqref{eq:taucosmogeneral}, followed
by division by $f_{\rm BH}$, yields Eq.~\eqref{eq:populationconvolution}
term by term.  Dimensionally, the line element, number density, and cross
section contribute length, inverse volume, and area, so the result is
dimensionless.  No luminosity-distance factor appears: optical depth is the
fractional survival exponent along the ray, and geometric dilution affects the
absorbed and unabsorbed intensities identically.  Cosmological redshift enters
only through the proper path element, the physical density, and the local
frequency shown explicitly above.

Equation~\eqref{eq:physicalnumberdensity} is an assumption about population
history, not a geometric identity.  An evolving population is included by
multiplying its right-hand side by an assembly function
$\mathcal{E}(M,z)$ with $\mathcal{E}(M,0)=1$.  We set
$\mathcal{E}=1$ in the plotted benchmarks.  This is appropriate for primordial
objects conserved after formation and is deliberately generous for stellar
remnants and massive black holes, which assemble with time; realistic assembly
histories generally reduce their integrated depths.

\clearpage

\section{AAA pole-extraction stability}
\label{app:aaapole}

The AAA audit uses the complex reflection amplitude, not the smoothed phase or
its numerical derivative.  We vary the fitted interval among
$[0.20,0.56]$, $[0.22,0.54]$, $[0.24,0.52]$, $[0.20,0.52]$, and
$[0.24,0.56]$; retain every second through every fifth point; and use relative
tolerances $10^{-4}$, $3\times10^{-5}$, and $10^{-5}$.  The algorithm's
small-residue cleanup is enabled.  For each fit the reported candidate is the
largest-residue pole in the predeclared physical search box stated in
Sec.~\ref{sec:numerics}; no nearest-known-pole criterion is used.

\begin{figure}[ht]
 \centering
 \includegraphics[width=0.88\linewidth]{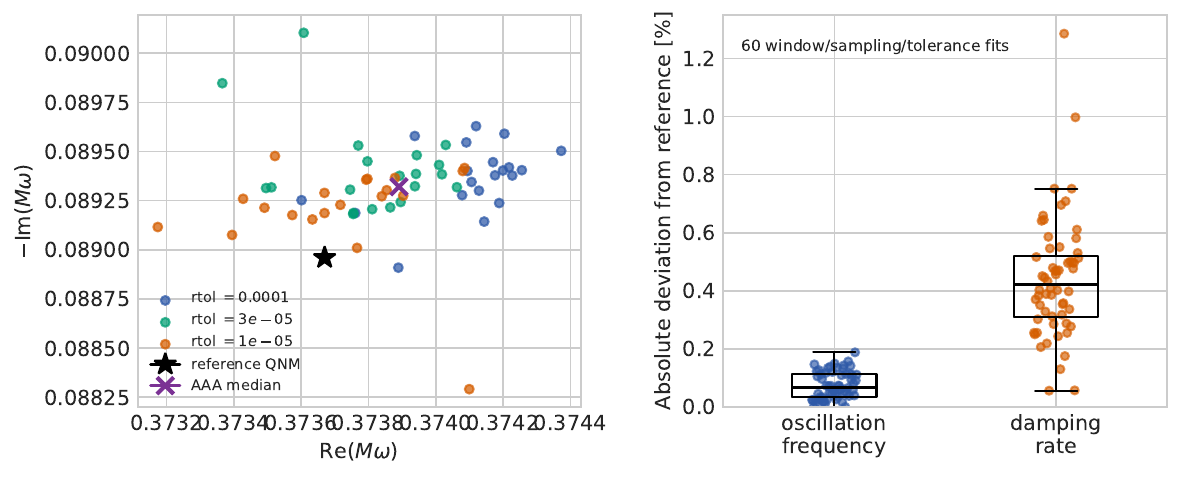}
 \caption{AAA extraction of the fundamental quadrupolar Schwarzschild pole.
 Left: selected poles for 60 frequency-window, sampling, and tolerance choices;
 the star is $0.37367-0.08896i$ and the cross is the AAA median.  Right:
 fractional deviations of the real and damping components from that reference.
 The pole is selected by residue within a fixed broad search box, not by
 proximity to the reference value.}
 \label{fig:aaapole}
\end{figure}

Figure~\ref{fig:aaapole} shows that the extracted pole forms one compact cluster
around the reference Schwarzschild fundamental.  The damping rate is less
stable than the oscillation frequency, as expected for continuation away from
a finite real interval, but its worst deviation remains $1.29\%$.  Spurious
poles elsewhere in the rational approximants move strongly with window and
tolerance and have much smaller residues; they are not interpreted as QNMs.

\bibliographystyle{JHEP}
\bibliography{biblio}

\end{document}